\documentclass[useAMS,usenatbib]{mnras}

\usepackage{tabto}
\usepackage{float}
\usepackage{color}
\usepackage{amssymb} 
\usepackage{amsfonts,amsmath,amscd} 
\usepackage[subnum]{cases}
\usepackage{mathtools}
\usepackage{verbatim}
\usepackage{t1enc}
\usepackage[T1]{fontenc}
\usepackage{ae,aecompl}
\usepackage[dvips]{graphics}
\usepackage{adjustbox}
\usepackage{eufrak} 
\usepackage{euscript} 
\usepackage{graphicx} 
\usepackage[normalem]{ulem} 
\usepackage{makeidx} 
\usepackage[usenames,dvipsnames]{pstricks}
\usepackage{epsfig}
\usepackage{pst-grad} 
\usepackage{pst-plot} 
\usepackage{pstcol,times,graphics,pstricks,pst-node,pst-coil} 
\usepackage{multicol} 
\usepackage{multirow} 
\usepackage{indentfirst} 
\usepackage{natbib} 
\usepackage{setspace} 
\usepackage{threeparttable}

\newcommand{\specialcell}[2][c]{%
  \begin{tabular}[#1]{@{}c@{}}#2\end{tabular}}

\title[CVs in MOCCA GCs -- present-day population]
{MOCCA-SURVEY database I. Accreting white dwarf binary systems in globular clusters -- I. cataclysmic variables -- present-day population}
\author[Belloni et al.]{Diogo Belloni$^{1,2}$\thanks{E-mail: belloni@camk.edu.pl (DTB)}
Mirek Giersz$^1$, Abbas Askar$^{1}$, Nathan Leigh$^{3}$ \newauthor Arkadiusz Hypki$^{4}$\\
$^{1}$ Nicolaus Copernicus Astronomical Centre, Polish Academy of Sciences, ul. Bartycka 18, PL-00-716 Warsaw, Poland \\
$^{2}$ CAPES Foundation, Ministry of Education of Brazil, DF 70040-020, Brasilia, Brazil \\
$^{3}$ Department of Astrophysics, American Museum of Natural History, Central Park West and 79th Street, New York, NY 10024, USA \\
$^{4}$ Leiden Observatory, Leiden University, PO Box 9513, NL-2300 RA Leiden, the Netherlands}
\begin{document}

\date{Accepted 2016 July 25. Received 2016 July 22; in original form 2016 April 21}

\pagerange{\pageref{firstpage}--\pageref{lastpage}} \pubyear{2016}

\maketitle

\label{firstpage}

\begin{abstract}
In this paper, which is the first in a series of papers associated with cataclysmic
variables and related objects, we introduce the CATUABA code,
a numerical machinery written for analysis of the MOCCA simulations,
and show some first results by investigating the present-day population
of cataclysmic variables in globular clusters.
Emphasis was given on their properties and the observational selection effects when observing and
detecting them. In this work we analysed in this work six models, including three with Kroupa 
distributions of the initial binaries. We found that 
for models with Kroupa initial distributions, considering the standard
value of the efficiency of the common envelope phase adopted
in BSE, no single cataclysmic variable was formed only via binary 
stellar evolution, i.e. in order to form them, 
strong dynamical interactions have to take place.
We show and explain why this is inconsistent with
observational and theoretical results. Our results indicate that the population of 
cataclysmic variables in globular clusters is, mainly, in the last stage of
their evolution and observational selection effects can 
drastically change the expected number of observed cataclysmic variables. We show that the probability of
observing them during the outbursts is extremely small and conclude that the best way of
looking for cataclysmic variables in globular clusters is by searching for variabilities 
during quiescence, instead of during outbursts. For that, one would need a very
deep observation which could reach magnitudes $\gtrsim 27$ mag. 
Finally, we argue that cataclysmic
variables in globular clusters are not necessarily magnetic.
\end{abstract}


\section{INTRODUCTION}
\label{introduction}

The study of star clusters plays an important role in our 
understanding of the Universe since these systems are natural 
laboratories for testing theories of stellar dynamics and evolution.
Particularly, globular clusters (GCs) are one of the most 
important objects for studying the formation and the 
physical nature of exotic objects such as X-ray binaries, 
degenerate binaries, black holes and blue straggler
stars, which in turn provides basic information and 
tools that can help us to understand the formation and 
evolution processes of star clusters, galaxies and, in general,
the young Universe. 

Among the most interesting objects in GCs 
are the cataclysmic variables (CVs) that are interacting
binaries composed of a white dwarf (WD) that accretes matter
stably from a main-sequence (MS) star -- or, in the last 
stage of their evolution, a brown dwarf (BD) -- 
\citep[BD; e.g.][for a comprehensive review]{Knigge_2011_OK}. 
CVs are subdivided according to their photometric 
behaviour as well as the magnetic field strength of the WD,
being, mainly, magnetic CVs (where the accretion is partially 
or directly via magnetic field lines) and non-magnetic CVs
(where the accretion is via an accretion disc). Among the 
non-magnetic CVs, the most prominent subgroup is that composed
of dwarf novae (DNe) which exhibit repetitive outbursts due to 
the thermal instability in the accretion disc 
\citep[e.g.][for a review]{Lasota_2001}.

\subsection{CV formation}
\label{cv_formation}

CVs are believed to form from an MS-MS binary that
undergoes a common envelope phase (CEP) when
the more massive MS evolves to a giant \citep{Paczynski_1976}.
In such CEP, the dense stellar giant core and the MS spiral
towards each other with the expansion and loss of the common envelope.
Most of the angular momentum is lost with the envelope which
leads to an orbital period orders of magnitude shorter
\citep[e.g.][]{Webbink_1984}.
After the CEP, a pre-CV is formed in a detached WD-MS binary.
Because of angular momentum loss (see below), the separation
between the stars decreases up to the formation of
the CV, when the MS starts filling its Roche lobe.

In order to form a CV through the above-mentioned scenario, 
the initial MS-MS binary should approximately have the 
following properties: (i) the more massive star has 
$M \lesssim \; 10 \; {\rm M}_\odot$; (ii) the less massive star 
is a low-mass MS; (iii) the mass ratio is $\lesssim$ 0.25;
(iv) the separation has to be sufficient to allow the primary
to expand to a point where it could form a degenerate core 
(pre-WD) and to permit the contact 
(the CEP).

The reasons for these conditions come almost naturally. The mass of the more massive
star has to be in the acceptable range to form a WD. This justifies the first condition.
Besides, the mass ratio after the CEP has to be $\lesssim$ 1
 \citep[e.g.][]{Hellier_2001}.
Otherwise, the mass transfer would
be unstable and such instability would precipitate further
mass transfer and then a merger would occur. This is 
because the mass transfer rate becomes very large and the WD cannot 
steadily burn the accreted material, then it swells 
up to become a giant, producing a common envelope 
binary and a merger of the stars. The WD mass 
cannot be greater than $\sim 1.44 \; {\rm M}_\odot$; thus, in order
to have the mass ratio less than one, the secondary mass cannot 
be larger than the WD mass, after the CEP. This
justifies the second condition. About the mass ratio, if the initial 
mass ratio is great (q $\sim 1$),
then both stars will have similar evolutionary time-scales. In this way,
the stars would become WDs at roughly the same time. For a CV however,
 one star has to be a WD and the other one an MS. This explains
the reason for the third condition. Finally, if the separation was too 
large, the more massive star
could not fill its Roche lobe. While if it was too small, the Roche
lobe overfilling could lead to a merger. Then, the separation
has to be ideal to allow for both the formation of a WD-like core and 
the close post common envelope binary.

\subsection{CV evolution}
\label{cv_evolution}

Non-magnetic CVs are usually separated in the following way,
with respect to the orbital period: (i) if the donor is an MS
and $P_{\rm orb} \gtrsim $  3 h, then it is called a long-period
CV; (ii) if the donor is an MS and 
$P_{\rm min} \lesssim P_{\rm orb} \lesssim $ 2 h, 
then it is called a short-period CV; and (iii) if the donor
is a BD, it is called a period bouncer CV (in this case,
$ P_{\rm orb} \gtrsim  P_{\rm min}$).
Besides, in CVs (especially non-magnetic ones), 
angular momentum loss is the driving mechanism for
their long-term evolution. In the so-called standard model 
of CV evolution, the dominant angular momentum loss mechanism 
in long-period
CVs is magnetic braking \citep{Rappaport_1982}, 
whereas in short-period and period bouncer CVs
the driving mechanism is associated with the emission of 
gravitational radiation \citep{Paczynski_1967}.

Basically, two important features observed in non-magnetic CVs 
need to be explained by the standard model. First, the absence 
of systems in the range of 2 h $\lesssim P_{\rm orb} \lesssim $  3 h,
known as the period gap, \citep[e.g.][and references therein]{Zorotovic_2016}
and, secondly, the existence of a 
period minimum $P_{\rm min} \approx$ 82 min \citep[e.g.][]{Gansicke_2009}.
The standard model reasonably fulfils its role in explaining the
observational properties of the CVs.

Briefly, the standard model can be summarized as follows: after the birth
of the CV, it will evolve towards short periods due to angular momentum loss. When it reaches
$P_{\rm orb} \sim$ 3 h (upper edge of the period gap),
the donor becomes fully convective and \citet{Rappaport_1983} 
proposed that at this point the magnetic braking turns off or becomes less efficient
(disrupted magnetic braking scenario). This results in a decrease of the
mass transfer rate, which allows the donor star to reestablish its equilibrium
and to stop filling its Roche lobe. Then, the system becomes
detached since the mass transfer stops. However, such a
detached WD-MS binary keeps evolving towards shorter periods 
due to gravitational radiation. When $P_{\rm orb} \sim$ 2 h (inner edge of the period gap),
the Roche lobe has shrunk enough to restart mass transfer and the
system becomes a CV again. After that, at some point during its evolution
($P_{\rm orb} \sim P_{\rm min}$), the mass loss rate from the secondary drives it 
increasingly out of thermal equilibrium until the thermal time-scale exceeds
the mass loss time-scale and it expands in response to the mass loss, thus, 
increasing $P_{\rm orb}$. Consequently, a large number of CVs are
 expected to be near the period minimum (known as the period spike)
or be in post-period minimum phase -- indeed, the abundance ratio
for long-period CVs,  short-period CVs and period bouncers,
respectively, is roughly 1:30:70 \citep[e.g.][]{Howell_1997}.

\subsection{CV observation}
\label{cv_observation}

Since the birth of interest in CVs \citep[e.g.]
[for a historical review]{Warner_1995_OK}, several breakthroughs
 have been taking place in the field, especially due to the 
Sloan Digital Sky Survey which has provided a reasonable
sample that reaches 
deeper magnitudes and which is capable of recognizing  
very faint CVs near and beyond the period minimum 
($P_{\rm min}$). 
Such breakthroughs include: the confirmation of
the disrupted angular momentum loss at the period gap \citep{Zorotovic_2016};
the discovery of period bouncers with BD donors \citep{Littlefair_2006};
the discovery of the period spike around $P_{\rm min}$ \citep{Gansicke_2009};
among others. They allowed the community
to considerably improve the observational data to confront 
theoretical predictions which, in turn, have led to a stronger
corroboration  between theory and observation.

All the previous discussion, theoretical and observational, are mainly
concerned with CVs in the field. For CVs in GCs, the same is not always true
because of the influence of dynamical interactions,
the ages and distances of the GCs, and the corresponding observational
selection effects.
Some observational efforts have also
 been made regarding CVs in GCs, especially a search for an optical 
counterpart of {\it Chandra} X-ray observations 
\citep[e.g.][for a review on CVs in GCs]{Knigge_2012MMSAI}. 

In general, there are four main approaches that have been used
for detecting CVs in GCs. In what follows, we will describe them
briefly including also the few important works associated
with them.

\subsubsection{Variability during outbursts}
\label{cv_observation_variability}

In all possible sorts of variability related to CVs, the most
explored is that regarding the DN outbursts which
last from days to tens of days and result in an increase of
luminosity by roughly 2 -- 5 mag. 

The first major investigation of CVs in GCs was done by \citet{Shara_1996},
who analysed 12 epochs of {\it Hubble Space Telescope (HST)} observations 
of the GC 47 Tuc and recognized only one DN and one DN candidate.
Other DNe have been discovered through variability during outbursts
after that in different clusters 
\citep[e.g.][]{Shara_2005,Knigge_2003,Servillat_2011}.
In a survey-like search for DNe, \citet{Pietrukowicz_2008} analysed 16 Galactic 
GCs and yielded two new certain DNe: M55-CV1 and M22-CV2. They found that the 
total number of known DNe in the Galactic GCs is 12 DNe, distributed among 
seven clusters.

Some comments are needed at this point. As \citet[][Section 5.5]{Knigge_2012MMSAI} 
has already pointed out, both \citet{Shara_1996}
and \citet{Pietrukowicz_2008} concluded that DNe are rare
in GCs based on the properties of observed CVs in the Galactic
field, which seems to be a biased sample of the
real population of CVs in the field. If most CVs in the field
are, in fact, period bouncers, then the observed CV population
in the field (especially the bright ones) is not representative 
of the real population of CVs in the Galactic field. 
This is mainly due to the very small duty cycle
(fraction of the DN cycle that a DN is in outburst)
associated with period bouncers. Hence, in this case, 
most CVs in the field are unobservable at any given time,
and a significant population of hidden CVs exists.
Thus, the conclusion that DNe are rare in GCs is not 
necessarily correct, since it could just be that they 
are hard and/or unlikely to observe, since there are 
more period bouncer CVs than originally expected
\citep{Knigge_2012MMSAI}.

Therefore, it turns out that identifying CVs through
their variability during outbursts is unlikely to 
reveal the intrinsic population of CVs in GCs since
one should be very lucky to detect the outbursts
in a sequence of epochs from {\it HST}, given the extremely
small duty cycle associated with the CVs in GCs.

\subsubsection{H$\alpha$ imaging}
\label{cv_observation_haplha}

Another possible way to detect CV candidates is
using H$\alpha$ imaging \citep[e.g.][]{Cool_1995},
since systems that exhibit an excess in H$\alpha$ show
evidence of variability. This technique is generally
used to study the counterparts of X-ray sources and
has revealed few CVs in some GCs 
\citep[e.g.][]{Grindlay_2006,Pietrukowicz_2009}.
However, doubt remains about the deepness
of observations using these techniques, i.e.
if they are able to detect the faint population
of CVs in GCs. 

So far, \citet{Cohn_2010} have reached
magnitudes as deep as 28 for H$\alpha$ and 26 from
optical observations of the {\it HST}. 
Their study seems to be
the least affected by this kind of bias and will
be used in this paper as the object for comparison
with our results.

\subsubsection{FUV band with {\it HST}}
\label{cv_observation_fuv}

Another way to detect potential CVs is through their
colours. CVs tend to be bluer due to accretion
processes. In fact, the energy released from this 
mechanism makes the region close to the WD hotter, which
in turn, makes the CVs bluer than typical stars in GCs.

This implies that looking for them in the far-ultraviolet
(FUV) band with {\it HST} is a good way to find
CV candidates \citep[e.g.][]{Dieball_2010}. 
Especially because most MSs in GCs
emit in infrared which nulls the problem with crowding.

\citet{Dieball_2010} carried out a detailed search in the
core of M80 and found few candidates. However, due to their 
limiting magnitude, they could only detect bright CV candidates.
In this way, the detection of the faint CVs using the {\it HST}'s FUV
detectors might also be problematic.

\subsubsection{X-rays}
\label{cv_observation_fuv}

The high resolution that has been achieved with {\it Chandra} 
allows us to, in fact, reach binaries with compact objects
in GCs, especially in their cores.

With regards to CVs, many GCs have been studied with
{\it Chandra} down to $\sim 10^{32}$ erg s$^{-1}$ \citep[e.g.][]{Pooley_2010}. 
Additionally, the identification of optical counterparts with deep
{\it HST} imaging has allowed for the recognition of many CV candidates
\citep[e.g.][]{Bassa_2004,Huang_2010,Cohn_2010},
although the number of such candidates is far from
the predicted number of CVs in the observed clusters.

Finally, it is valid to note that below $\sim 10^{32}$ erg s$^{-1}$,
a large fraction of X-ray sources do not have secure optical
counterparts. Below this value, the sources can be chromospherically
active binaries (or near-contact binaries of MSs), CVs, foreground 
and background objects, quiescent low-mass X-ray binaries, millisecond 
pulsars or black hole X-ray binaries. Any conclusions drawn from a 
comparison between the results of our simulations and observations of 
CVs with small X-ray luminosities should be taken with a grain of salt.  
This is because the observational sample can be regarded as something of 
an upper limit, due to an increased probability of contamination from 
active binaries, chromospherically active stars, accreting neutron stars 
and black holes, etc.

\subsubsection{Classical novae}
\label{cv_observation_cn}

On the subject of classical novae (CVs with high mass transfer rates
and stable and hot discs), it is worth mentioning some
observational evidence of different nova rates with respect
to the Galactic field.

\citet{Curtin_2015} 
detected novae in a survey of GCs in three Virgo 
elliptical galaxies (M87, M49 and M84). Such
a survey should not detect any novae if there were 
no enhancement of the nova rate due to dynamics. 

A similar result was reached by \citet{Shara_2004} while 
investigating one GC of M87. They concluded that
classical nova eruptions in GCs are up to 100 times 
more common than current detections in the Milky Way 
suggest.

This implies that dynamics are extremely
important in enhancing the detection rate of 
novae in GCs.

\subsubsection{What is the lesson from the observations?}
\label{lesson}

Given the crowding of GCs and the faintness of the intrinsic
population of CVs, confirming spectroscopically the many CV 
candidates that have been observed as real CVs  is 
challenging \citep[e.g.][]{Knigge_2003,Thomson_2012}. 
On the other hand, the usage of a combination of different 
techniques (H$\alpha$ and FUV imaging, X-ray, colour, etc.) 
can provide almost guarantees the confirmation of CVs, especially for DNe. 
For instance, \citet{Cohn_2010} used H$\alpha$ imaging
and colours to infer that some {\it Chandra} X-ray sources are
CVs.

As it seems, a combination of techniques can provide us
with the potential number of CVs in GCs. The only problem is whether
or not we can reach faint CVs, given the observational
limitations and biases of each technique when combined together.

This indicates that the biases and observational limitations can 
lead us to incorrect impressions about the nature of CVs in GCs.

\subsection{Nature of CVs in GCs}
\label{cv_nature}

In the most recent survey-like search for DNe, 
\citet{Pietrukowicz_2008} conclude that `ordinary DNe are
 indeed very rare in GC'. Then, either the predicted number of CVs is not correct,
or the predicted CVs would be non-DNe. Nevertheless,
such observational findings do not corroborate theoretical predictions.

First, theoretical works predict that most CVs should be DNe 
\citep[e.g.][]{Knigge_2011_OK}. Secondly, some previous numerical 
investigations regarding CVs in evolving GCs estimate around 100-200 CVs
in massive GCs. In the most recent study on the matter, \citet{Ivanova_2006} 
predicted that around 200 CVs would be present in a typical massive GC. 
Besides, such CVs would have different properties from the CVs in the field.
For example, only $\sim$ 25 per cent of CVs were formed in binaries that
would become CVs in the field. Also, approximately 60 per cent of the CVs
did not form via a CEP.

This corresponds to a rather strong inconsistency between 
theory and observation and the most popular hypothesis that 
attempts  to explain the so-called absence of DNe in GCs is based 
on the mass transfer rate and the WD magnetic field. 
\citet{Dobrotka_2006} proposed, using the disc instability model 
(DIM), that a low mass transfer rate combined with a moderately 
strong WD magnetic field can disrupt the inner part of the disc, 
preventing, in turn, an outburst in such CVs.

In the mid-1990s, the community started to think that CVs in GCs
tend to be magnetic due to the work by \citet{Grindlay_1995} who
analysed three CVs in NGC 6397 and found the existence of He II line 
in such CVs. Also, \citet{Edmonds_2003a,Edmonds_2003b} argued in the same
direction in their studies of 47 Tuc.

The big problem with this argument is that not only magnetic CVs exhibit 
the He II line in their  spectra, but also other types 
of CVs \citep{Knigge_2012MMSAI}.  For instance,
\citet{Shara_2005} found  DN-like eruptions in CV2 
and CV3 in NGC 6397 which exhibited helium emission in 
their spectra. Therefore, it seems that such evidence is not
strong enough to claim that CVs in GCs are,
principally, magnetic ones.

The main reason for this suspicion is the attempt to explain
the discrepancies between observed CVs in GCs and those in the field.
This is because a WD magnetic field can prevent
instability in the disc and, in turn, suppress the occurrence of outbursts 
\citep{Dobrotka_2006}. Besides, it could explain to some extent,
the unique X-ray and optical properties found for the CVs in GCs.

However, the CV samples in GCs tend to be X-ray selected \citep{Heinke_2008}
which, in turn, favours the detection of magnetic CVs (brighter in X-ray
than the non-magnetic counterpart). Unfortunately, few investigations
went deep enough to detect the non-magnetic CVs ($\lesssim 10^{30}$ erg s$^{-1}$)
in the X-ray and more efforts should be put in this direction.

Another point in favour of the idea that CVs in GCs are magnetic
comes from the fact that magnetic CVs
are usually associated with massive WDs \citep{Ivanova_2006}. In fact,
in GCs, the dynamically formed CVs tend to have higher WD masses which,
in turn, favours the hypothesis of magnetic CVs.

Above all, such a hypothesis is not well established and can be contested. 
As has already been mentioned, most CVs should be DNe 
\citep{Knigge_2011_OK}. Besides, not only magnetic CVs
produce the above-mentioned He II emission.
Over and above, many optical counterparts of X-ray
sources have been recognized as reasonable CV candidates
in some GCs \citep[e.g.][]{Cohn_2010}, although such numbers
are still far from the predicted number of CVs.

\subsection{Structure of the paper}
\label{paper_structure}

From the last subsection, we can say that
the `CV problem in GCs' is not yet well understood and solved.
In order to contribute to such a discussion,
this is the first paper of a series that concentrates on CVs and related 
objects (such as AM CVn and symbiotic stars) in evolving GCs that attempt 
to correlate CV properties (and also AM CVn and symbiotic star properties)
with cluster initial and present-day parameters.

The main objective of this first paper is to introduce the CATUABA code
(Section \ref{catuaba}) that will be used in this series of papers to derive properties 
of CVs and related objects from MOCCA simulations. In order to test its efficiency
and coherence, we concentrate on only six models with different initial conditions 
and different properties at the present-day (described in Section \ref{models}).
The models were simulated by the MOCCA code (Section \ref{mocca}) that simulates
real GCs on a time-scale of one day. Its speed allows for the simulation of hundreds of 
models in a short time and, in turn, permits a very detailed statistical analysis
of particular objects (like CVs) and their correlations with the cluster parameters.
The aim of this series of papers is to analyse the MOCCA database with respect
to CVs and related objects.

For convenience, we decided to separate the results
of this initial work into two separate papers. In this paper
we will concentrate mainly on the present-day (considered here
as 12 Gyr) population of CVs
present in the analysed models. In the next, we will discuss
mainly the CV progenitors and the main 
formation channels as well as their properties at the moment they are 
formed and the subsequent evolution up to present day. We will also deal
with more general issues like unstable systems, escaping binaries
that become CVs, and so on.

In Section \ref{mocca}, we describe the MOCCA and BSE codes. We also
make some comments with regards to the comparison between MOCCA and N-body codes. 
We describe the CATUABA code in Section \ref{catuaba}, and end the Section
by summarizing its main features.
The models used in this first work will be described in Section \ref{models}, and
in Section \ref{results}, the main results of the preliminary investigation
are presented. We show results associated with the clusters and their present-day
populations (PDPs) of CVs as well as results related to observational 
selection effects when searching for CVs in GCs. 
We also address some connections between our results and
observations and, also, between our investigation and previous studies.
We conclude and discuss the main
implications of these first results in Section \ref{conclusion}.
Finally, throughout this paper, we use some new
abbreviations, and for convenience, in Appendix \ref{ap1},
we clearly define all abbreviations in order to allow
the reader to consult them if necessary.

\section{MOCCA CODE}
\label{mocca}

In this section, we describe the MOCCA code that was used to 
simulate the cluster evolution. We also describe the BSE code that
is utilized by MOCCA to perform the stellar evolution. We end this
section by briefly comparing MOCCA's performance with that of N-body codes.

The MOCCA code \citep[][and references therein]{Hypki_2013,Giersz_2013}
is based on the orbit-averaged Monte Carlo technique
 for cluster  evolution developed by \citet{Henon_1971}, which 
was further improved by  \citet{Stodolkiewicz_1986}. It also includes the 
FEWBODY code, developed by \citet{Fregeau_2004}, to perform 
numerical scattering experiments of gravitational interactions.
To model the Galactic potential, MOCCA assumes a point-mass with 
total mass equal to the enclosed Galaxy mass at the galactocentric 
radius. The description of escape/capture processes in tidally limited 
clusters follows the procedure derived by 
\citet{Fukushige_2000}. The stellar evolution is implemented via
the SSE code developed by \citet{Hurley_2000} for single stars and
the BSE code developed by \citet{Hurley_2002} for binary evolution.

\subsection{BSE code}
\label{bse_code}

The most important part of MOCCA with regards to CVs and related
objects is the BSE code \citep{Hurley_2002}.
BSE models angular momentum loss mechanisms, although its implementation is
not up-to-date \citep[][for a revised model]{Knigge_2011_OK}. 
Mass transfer occurs if either star fills its Roche lobe and 
may proceed on a nuclear, thermal, or dynamical time-scale. 
Prescriptions to determine the type and rate of mass
transfer and the response of the primary to accretion 
are implemented in BSE. The overall CV evolution can
be recovered by BSE, although some comments are necessary
here.

With respect to the CEP, BSE assumes that the common envelope
binding energy is that of the giant(s) envelope involved
in the process. In order to reconcile the prescription
developed by \citet{Iben_Livio_1993}, we use the recommended value for 
the CEP efficiency which is 3.0. This is greater than the usually
adopted values, but we kept it for the initial investigation.

With respect to CV evolution, the donor radii for long-period
CVs are not  increased in BSE. There is observational evidence of
an increase in the radii of CV donors with radiative
cores \citep[][see their fig. 6]{Knigge_2011_OK}. 
Besides, the mass transfer rate is not adjusted
to be in agreement with those derived from observations
in the field. This implies a different CV evolution, which leads to the absence
of the period gap and to a different period minimum. 
Additionally, for period bouncers, the mass-radius relation
leads to a faster increase in the period after the period minimum.

Even though some efforts should be put into the improvement
of BSE in this direction, for the purposes
of this work, the approach already implemented in BSE seems reasonable. Furthermore,
we use standard values for all parameters in the BSE code \citep{Hurley_2002}
Nonetheless, we have to bear in mind these factors while analysing 
the results.

\subsection{Comparison with N-body codes}
\label{nbody_codes}

MOCCA was extensively tested against N-body codes. 
For instance, \citet{Giersz_2013} 
concluded that MOCCA is 
capable of reproducing N-body results with reasonable precision, not 
only for the rate of cluster evolution and the cluster mass distribution, 
but also for the detailed distributions of mass and binding energy of 
binaries. Additionally, \citet{Wang_2016} also compared MOCCA with
the state-of-the-art NBODY6++GPU and showed good agreement
between the two codes.

In general, many of the simplifying assumptions adopted in the Monte 
Carlo method that would be naturally accounted for in an N-body code 
are unimportant in the regime of cluster masses where Monte Carlo models 
are ideally suited.  For example, Monte Carlo methods treat both binary 
evolution and direct single-binary and binary-binary encounters as isolated
 processes, with no chance of being interrupted due to a dynamical encounter.  
This was recently shown to be a valid assumption in clusters more massive than 
$\gtrsim$ 10$^5$ M$_{\odot}$, with the probability of interruption being of 
the order of a per cent or less \citep{Geller_2015,Leigh_2016}. The approximations 
underlying the Monte Carlo method are in many ways perfectly suited to model
massive cluster evolution, the regime of star cluster evolution where N-body 
models cannot typically go (at least not without being severely limited by the 
computational expense).

In essence, MOCCA is ideal for performing big surveys and for 
carrying out detailed studies of different types of objects like CVs
(this paper), blue straggler stars \citep[][]{Hypki_2013,Hypki_2016a,Hypki_2016b}, 
intermediate-mass black holes \citep[][]{Giersz_2015}, X-ray binaries, etc., and 
provides good agreement with N-body codes.

\subsection{Why MOCCA?}
\label{nbody_codes}

The MOCCA code has been developed for more than 20 years 
\citep{Giersz_1998,Giersz_2001,Giersz_2006,Giersz_2008}, and its current version
 \citep{Hypki_2013,Giersz_2013} is characterized by its high speed, its modularity 
and its detailed information about each and every object in the system.\footnote{The 
simulations were performed on a PSK cluster at the Nicolaus Copernicus 
Astronomical Centre in Poland.} 

In this way, MOCCA is convenient for two purposes that will help with the
investigation of CVs and related objects in GCs: (i) its high speed allows for
generating a big database covering a huge range in the 
cluster parameter space and subsequently allows for powerful statistical
analysis regarding the cluster parameters and the studied objects;
(ii) its list of the most relevant events during the life of each star
in the cluster admits a very detailed investigation concerning 
formation/destruction channels, strength of dynamical interactions,
and so on.

To sum up, the MOCCA code is ideal for performing big surveys and 
for carrying out detailed studies of different types of exotic objects
like CVs, blue straggler stars, degenerate binaries, etc.

\subsection{The MOCCA-SURVEY Database}
\label{bigsurvey}

\citet[][see their table 1]{Askar_2016b} describes the set of 1950 GC models (called MOCCA-SURVEY) 
that were  simulated using the MOCCA code. The models have quite diverse parameters 
describing not only the initial global cluster properties, but also star and binary parameters. 

The clusters vary with respect to:
(i) metallicity: 0.02, 0.006, 0.005, 0.001 and 0.0002;
(ii) binary fraction: 0.95, 0.3, 0.1 and 0.05;
(iii) King model parameter (${\rm W_0}$): 9.0, 6.0 and 3.0;
(iv) tidal radius: 120.0, 60.0 and 30.0;
(v) cluster concentration as measured by the ratio between 
tidal and half-mass radii: 50.0, 25.0 and tidally filling;
(vi) initial binary properties (period, eccentricity and 
mass ratio), being the initial mass function (IMF) given by \citet{Kroupa_2001};
(vii) supernova natal kicks for black holes distributed 
according to \citet{Hobbs_2005} or reduced according to 
mass fallback description given by \citet{Belczynski_2002}.

Despite the fact that the models in the MOCCA-SURVEY were not selected to match the observed Milky Way GCs,
they agree well with the observational properties of the observed GCs
\citep[][see their fig. 1]{Askar_2016b}. They conclude that the MOCCA-SURVEY cluster 
models are representative for the Milky Way GC population. The six models discussed in the 
paper were chosen from the MOCCA-SURVEY database.

\section{CATUABA code}
\label{catuaba}

In this section we describe the code CATUABA\footnote{Catuaba
 is a vigorous-growing, small tree that produces yellow and orange
 flowers and small, dark yellow, oval-shaped, inedible fruit.
It grows in the northern part of Brazil, the Amazon, Para, Pernambuco,
 Bahia, Maranhao, and Alagoas. Catuaba has origin in one Brazilian 
indigenous language (tupi-guarani) and means `strong plant' since
its bark tea is a central nervous system stimulant and also an 
innocent aphrodisiac.} ({\bf C}ode for {\bf A}nalysing and 
s{\bf TU}dying c{\bf A}taclysmic variables, sym{\bf B}iotic stars
 and {\bf A}M CVns), which is capable of investigating CVs. In future
improvements, it will be extended to also include symbiotic stars and
AM CVn.

First, we describe how the CATUABA code recognizes the PDP
in clusters simulated by MOCCA and also how it separates
the CVs according to their main formation mechanisms. Secondly,
we explain the main physical assumption included in the code
to study the observational properties of the PDP
of CVs in a simulated cluster. Thirdly, we describe other features included
in the code and finish the section with a summary of its operation based on
a flow chart of its prime provisions for the analysis of the CV population.

\subsection{CV populations}
\label{population}

In this subsection we describe the three populations used
to study the CVs. They correspond to the properties of the same CVs 
at three different times.

\subsubsection{The present-day population}
\label{pdp}

The first step in the construction of an inventory of CVs in a GC is 
the recognition of their population at the present-day age of such a cluster.
This is possible due to recurrent snapshots (around every 200 -- 250 Myr)
of the system recorded
by MOCCA during the simulation. We have chosen a snapshot of the cluster
 at around 12 Gyr to be the present-day cluster. Then, an extraction of the 
PDP of CVs is easy, based on the definition 
given in Section \ref{introduction}.

Once the stars in the PDP of CVs are recognized, a complete study of 
them is done from the progenitor population (see below)
 up to the PDP. MOCCA provides a full history of the dynamical and physical evolutionary events 
of all stars in the system \citep[][for blue straggler
stars]{Hypki_2013} through the MOCCA-MANAGER code.

\subsubsection{The progenitor and formation-age populations}
\label{pdp}

Given a star in the cluster, MOCCA-MANAGER creates a complete
history of all relevant steps during the life of the star thereof. In this
chronological list, all stellar evolution and dynamical events are recorded. Therefore,
we can easily get the properties of the progenitor population and
the formation-age population of CVs in the cluster. 
The progenitor population is defined
here as the population of all binaries that are progenitors of the PDP of CVs, 
i.e. the population of CVs at the birth of the cluster.

Now, the formation-age 
population is the population related to the birth of the CV itself, in the
sense that mass transfer starts from MS on to a WD. Obviously, for the
formation-age population, the time is not unique as in the progenitor population and the PDP. During the cluster 
evolution, each CV is formed at a specific time, different from the other ones.

One comment about dynamical exchanges and the progenitor population of CVs
is necessary at this point.

First, we define a dynamical 
exchange (or just exchange) by a process in which a binary
has one of its components replaced by another star in a binary-single
or a binary-binary interaction. It can happen (and happens) that
a CV can be formed due to exchanges, in the sense that both the 
CV components are not the same as those of the primordial binary.

Secondly, the progenitor population of CVs for the cases without exchange is easily determined,
since the components of the primordial binary are the same components of the CV.
Although the situation is not straightforward when there is exchange in the 
history of the CV. The initial properties in the case of exchange are
obtained by getting the properties of the binary with the smaller 
period (if the exchange took place in a binary-binary interaction) or 
the properties of the binary (if the exchange happened in a
single-binary interaction). In the case that both components
of the CV in the PDP were single stars at the cluster
birth, then such CVs are excluded from the progenitor population since there is no
initial binary associated. In summary:

(i) if both CV components are the same at the very beginning, then
the CV is formed from the primordial binary, i.e. there is no exchange during
the binary's life. Thus, this binary enters into the progenitor population;

(ii) if one CV component comes from one binary and the other
one comes from another binary, then the properties of the binary with 
the smaller size at the beginning is saved in the progenitor population;

(iii) if one CV component comes from a binary and the other one
was a single star at the very beginning, then the binary properties
are saved in the progenitor population;

(iv) if both CV components come from single stars, then there
is no binary at the very beginning associated with the CV. Since the progenitor population 
is defined for binaries only, we do not include any binary for such a 
CV in the progenitor population.

It is worthwhile to mention that only a few CVs enter in case (iv) above
(< 1 per cent for the models investigated here) which, in turn, will not 
change the conclusion derived from the analysis
of the progenitor population (in the second paper).

\subsection{Categorization of the present-day CVs}
\label{categorisation}

Given the complete history of events during the lives of
the CVs in the PDP, an attempt of categorizing them
comes naturally. Our choice was to separate them
 with respect to the importance of the dynamical
interactions in their lives. For that, we separate 
every CV in one of the following groups: (i) 
{\it binary stellar evolution group (BSE group)}: 
the formation of the CV and its subsequent evolution was 
{\it without} any influence of dynamical interactions.
This means that a CV in this group has formed from a 
binary in the progenitor population {\it only} through binary stellar evolution.
(ii) {\it weak dynamical interaction group (WDI group)}: 
the CV life was influenced only by {\it weak} dynamical
interactions, in the sense that the binding energy during
such interactions changes by a factor of less than 20 per cent
during the evolution of 
the binary in the progenitor population up to the formation of the CV and 
subsequently, up to PDP.
(iii) {\it strong dynamical interaction group (SDI group)}: 
the formation of the CV was influenced {\it strongly} by dynamical
interactions, in the sense that such interactions have
significantly changed the evolution of the binary in the progenitor population up to the
formation of the CV. After the formation of the CVs in this group,
there are no subsequent strong dynamical interactions.

Some comments are worthwhile to mention before proceeding further:

(i) Whether a CV belongs to the WDI group or the SDI
group is a priori arbitrary. It is based on the change of 
energy during the dynamical interactions. 
The cutoff between a weak dynamical interaction
and a strong one is 20 per cent of the initial binding energy. 
This cutoff is for only one dynamical interaction and
it is an arbitrary value connected to the average energy change
 according to Spitzer's formula for equal-mass systems 
\citep{Spitzer_BOOK}. Generally, for weak dynamical interactions,
most of the interactions are fly-bys, although they can also be resonant 
interactions, and for strong dynamical interactions, they can be 
resonant or exchange interactions.
As follows, if a binary undergoes one interaction and the change of its
binding energy is less than 20 per cent, this binary was involved in one weak dynamical
interaction; otherwise, it was subjected to one strong dynamical interaction.
If a binary underwent {\it only} weak dynamical interactions during its life, 
then this binary belongs to the WDI group. Now, if a binary
underwent {\it at least one} strong dynamical interaction in its life, then
this binary belongs to the SDI group.

(ii) The WDI group is formed by CVs which had only weak 
encounters during their lives. However, if the number
of these encounters is high for a CV in this group, 
their cumulative effect can be strong and the properties
of the progenitor binary can be strongly changed during its life.
Otherwise, if the cumulative effect of the weak dynamical
interactions is not strong, the CV will probably have
similar properties to those belonging to the BSE group.

(iii) The SDI group also includes binaries which
underwent either exchange or merger.

\subsection{CV properties}
\label{properties}

In this part of the methodology, we provide the main
assumptions related to the CV properties. We
state the main theoretical and empirical relations
used in CATUABA. With such properties, we can estimate
an upper limit for the probability of observing
a CV (Section \ref{proba}).

Once the PDP of the CVs is acquired, we can start to analyse
the properties of the population thereof. Fortunately,
the BSE code in MOCCA is capable of giving basic information
about binaries in snapshots of the cluster, like masses,
orbital elements, bolometric luminosities, radii, and magnitudes. 
Nonetheless, in the precise case of CVs, BSE is not able
to model the typical behaviours of sub-types of CVs, including
DIMs. These processes need to be covered in our analysis.

This is necessary because one of the aims of the CATUABA code is to calculate
the probability of observing CVs as a function of the GC distance and
the optical limiting magnitude. Besides,
it is rather important to have a model of how the brightness varies with time
for systems that undergo variability (like CVs). In the specific case
of CVs, the variability is caused by either eclipses or outbursts,
and the chances of detecting a CV depend on how its magnitude varies 
with time due to either eclipses or recurring outbursts. In order to compute 
such brightness variations due to outbursts, 
the well-accepted DIM will be used as it provides maximum
values for the CV luminosity during quiescence and during outburst.

\subsubsection{CV types}
\label{cv_type}

As we have already mentioned, one expects that
most CVs are DNe \citep{Knigge_2011_OK}.
In this sense, we should be able to model DNe. 

DNe are characterized by outbursts due to the thermal instability
in the accretion disc \citep[e.g.][]{Smak_2001}. During the outburst,
their luminosity increases by typically 2 -- 5 mag. Also, outbursts last
from days to tens of days, and occur every tens of days or even decades.

In order to classify a CV as a DN, we adapted equation A.3 and A.4 of
\citet{Lasota_2001} which defines the limits for the disc stability based
on the CV properties, the position inside the disc and the viscosity 
parameter $\alpha$. Hence, the mass transfer rate $\dot{M}_{\rm B}$ defines the 
limit between hot/ionized/stable and unstable disc, and the mass transfer rate
$\dot{M}_{\rm A}$ defines the limit between cold/neutral/stable and unstable disc. 
They are

\begin{multline}
\dot{M}_{\rm A} \ = \ 6.344 \times 10^{-11} \ \alpha_{\rm c}^{-0.004} \ \left( \frac{M_{\rm WD}}{\rm {\rm M_\odot}} \right)^{-0.88} \\ 
\times \left( \frac{r}{10^{10} \ {\rm cm}} \right)^{2.65} \ {\rm M_\odot \ yr^{-1}}\label{MA}
\end{multline}

\noindent
and

\begin{multline}
\dot{M}_{\rm B} \ = \ 1.507 \times 10^{-10} \ \alpha_{\rm h}^{0.01} \ \left( \frac{M_{\rm WD}}{\rm M_\odot} \right)^{-0.89} \\ 
\times \left( \frac{r}{10^{10} \ {\rm cm}} \right)^{2.68} \ {\rm M_\odot \ yr^{-1}}\label{MB}
\end{multline}

\noindent
where $M_{\rm WD}$ is the WD mass, $\alpha_{\rm c}$ is the usual viscosity 
parameter for the cold branch of the \citet{Shakura_1973}
 solution, adopted here as $0.01$,
$\alpha_{\rm h}$ is the usual viscosity parameter for the hot branch 
of the Shakura-Sunyaev solution, set here as $0.1$,
and $r$ is the radial distance from the centre of mass. 
At each $r$, $\dot{M}_{\rm A}$ is the critical mass transfer rate 
below which the disc is cold/neutral and stable and $\dot{M}_{\rm B}$
is the critical mass transfer rate above which the disc is 
hot/ionized and stable.

Both values ($\dot{M}_{\rm A}$ and $\dot{M}_{\rm B}$) are computed
based on analytic fits concerning 1D time-dependent numerical models of accretion discs, 
using an adaptive grid technique and an implicit numerical scheme, 
in which the disc size is allowed to vary with time \citep{Hameury_1998}

Given the mass transfer rate of a CV in the PDP ($\dot{M}_{\rm tr}$), 
we have three different regimes: (i) if $\dot{M}_{\rm tr} > \dot{M}_{\rm B}$ 
everywhere in the disc, the disc is stable and hot;
(ii) if $\dot{M}_{\rm tr} < \dot{M}_{\rm A}$ everywhere in the disc, 
the disc is stable and cold; (iii) if $\dot{M}_{\rm A} < \dot{M}_{\rm tr} < \dot{M}_{\rm B}$,
in some ring $r$ of the disc, the disc is unstable and undergoes 
repetitive outbursts.

\subsubsection{Mass transfer and accretion rates}
\label{mtr_acc}

Before moving on, we have to explain how the mass transfer and accretion rates are calculated,
since this information is not given by the BSE code in MOCCA. The {\it mass transfer
rate} here is defined as the mass loss rate of the donor, i.e.
the rate at which the accretion disc is fed (in the case of a non-magnetic CV
or an intermediate polar CV). On the other hand,
the {\it accretion rate} is defined as the rate at which the mass flows
through the accretion disc towards the WD.

It is worth noticing that in long-period CVs in which the donor
is an evolved star (sub-giant), the mass transfer can also be driven 
by the  nuclear expansion of the donor star,  e.g. AKO 9 in 47 Tuc
\citep[][]{Knigge_2003}. Nevertheless, based on our definition of CV 
(Section \ref{introduction}), we are mainly concerned with unevolved donors,
which is reasonable, since MS stars are more common in GCs than more 
evolved donors by orders of magnitude.

On how the mass transfer rate is computed, for simplicity,
let us consider only one CV. From the list of all relevant events during 
the CV life, two quantities are used to compute the mass transfer rate, 
namely: time and donor mass. For each timestep in the CV life, 
the mass transfer rate is computed based on the amount of mass lost by the donor
due to angular momentum loss during this timestep. 
Thus, the mass transfer rate calculated here corresponds to
a crude estimation. The accretion rate will be treated latter
in this section.

We assume, a priori, that the disc is not disrupted and
the inner edge of the disc is the WD radius. 
Besides, the outer edge of the disc during outbursts is assumed to 
be 90 per cent of the WD Roche lobe \citep{Smak_1999} which, in turn, is 
computed based on the equation derived by \citet{Eggleton_1983}.

With these values for the viscosity parameter and edges of the disc,
it is straightforward to distinguish between stable and unstable discs
by using equation \ref{MA} and \ref{MB}.

In order to calculate the observational properties of the CVs
with unstable discs, we need to find the associated accretion
rates ($\dot{M}_{\rm d}$) through the disc, specifically 
during outburst ($\dot{M}_{\rm dO}$) and 
during quiescence ($\dot{M}_{\rm dQ}$).

The BSE code inside MOCCA is capable of giving some information
about the binary itself. Nevertheless, information about the
other components (hot spot and disc) for a DN is missing. 
Therefore, we adopt maximum values for the accretion rate
during quiescence and outburst. For the accretion rate during
quiescence we simply assume $\dot{M}_{\rm dQ} = \dot{M}_{\rm A}$;
while for the outburst, we adopt $\dot{M}_{\rm dO} = \dot{M}(d)$,
where $\dot{M}(d)$ is obtained from equation 3.27 of \citet{Warner_1995_OK},
which is qualitatively similar to the expression derived by
\citet{Lasota_2001}.

We choose to use the maximum values during
quiescence and during outburst because we are interested in an
upper limit for the probability of observing a CV. A similar
approach for the accretion rate is used in the StarTrack 
code \citep{Belczynski_2008} which deals with X-ray binaries.

Concerning the disc radius for a DN in quiescence, 
we adopt a value of 50 per cent of the
WD Roche lobe for the outer edge \citep{Warner_1996}.
While for the outburst, as we have already mentioned,
we use a value of 90 per cent of the WD Roche lobe \citep{Smak_1999}.

\subsubsection{CV visual luminosity}
\label{luminosities}

The WD and donor bolometric luminosities
are provided by the BSE code. We only need
additional information about the hot spot and the disc. The 
bolometric luminosity of the hot spot $L_{\rm bHS}$ is given by
\citep[e.g.][p. 83]{Warner_1995_OK}

\begin{equation}
L_{\rm bHS} \ = \ \left( \frac{f}{8} \right) \ \frac{G \; M_{\rm WD} \; \dot{M}_{\rm tr} }{r_{\rm d}} \label{LHP}
\end{equation}

\noindent
where $f$ $\sim$ 1.0 is an efficiency factor, and the bolometric luminosity of the disc $L_{\rm bd}$ is given
by \citep{Paczynski_1980}

\begin{equation}
L_{\rm bd} \ = \ \frac{ G \; M_{\rm WD} \; \dot{M}_{\rm d} }{2 r_{\rm d} } 
\left[ 1 - 3 \left( \frac{ R_{\rm WD} } { r_{\rm d} } \right) + 
2 \left( \frac{ R_{\rm WD} }{ r_{\rm d} } \right)^{ \frac{3}{2} } \right] \label{LD}
\end{equation}

\noindent
where $r_{\rm d}$ is the outer edge of the disc and  $R_{\rm WD}$ is the WD radius.

Once the bolometric luminosities are computed,
CATUABA calculates the bolometric correction for the hot spot
and for the disc. The hot spot bolometric correction is computed assuming 
an ellipsoid, with the semi-axes $s_{\rm a} =$ 0.043 $a$
and $s_{\rm b} =$ 0.22 $s_{\rm a}$, where $a$ is the semi-major axis
\citep{Smak_1996}. Given the shape of the hot spot and using
the bolometric correction coefficients given by \citet{Torres_2010}, the code computes the 
hot spot temperature and also the visual luminosity and magnitude.

A similar procedure is done for the disc, although it is necessary to integrate
over the entire spatial extent of the disc.
This is done following the procedure by \citet{Paczynski_1980} and also
using the bolometric correction coefficients given by \citet{Torres_2010}.

\subsubsection{CV X-ray luminosity}
\label{properties}

Another important observational property calculated by the 
CATUABA code is the X-ray luminosity. This is computed based
on equation 4 derived by \citet{Patterson_1985}, in the band 
[0.5 -- 10 keV], for a slowly rotating WD:

\begin{equation}
L_{\rm X} \ = \ \varepsilon \; \frac{G \; M_{\rm WD} \; \dot{M}_{\rm dQ}}{2 \; R_{\rm WD}} \label{XRAY}
\end{equation}

\noindent
where the factor $\varepsilon = $ 0.5 is the correction for the 
fraction of the X-rays emitted inwards and absorbed by the WD.

\citet{Patterson_1985} proposed an explanation for the origin of the
X-ray luminosity in CVs. They showed that, at small accretion rates 
(during quiescence for DNe),
the boundary layer (region between the accretion disc and the WD)
is the likely site of most of  the  observed hard X-rays. This is because
at small accretion rates the boundary layer is thin. In the case of
high accretion rates (during outbursts for DNe), the boundary layer
is thick and consequently the observed hard X-rays do not come from this
region \citep{Eze_2015}. Actually, one of the open questions on this matter
is associated with the origin of the hard X-ray luminosity observed
for systems with high accretion rate \citep{Mukai_2014}.
As there is no consensus with respect to systems with high accretion rate,
equation \ref{XRAY} is used only for the quiescent stage (for DNe)
or for CVs with stable and cold discs. There is no X-ray luminosity
for the DNe during outbursts or CVs with stable and hot discs modelled
in CATUABA. That is the why in equation \ref{XRAY} the $\dot{M}_{\rm dQ}$
is given by its value in the boundary layer, assumed here as being
the WD radius.

\subsubsection{DN time-scales}
\label{time-scales}

Important time-scales for DNe are the recurrence time and
the duration of the outbursts. The recurrence time can be
defined as the interval between two outbursts and 
is computed here based on the observational fit
done by \citet{Patterson_2011}

\begin{equation}
t_{\rm rec} \ = \ 318 \; \left( \frac{ M_{\rm donor}/M_{\rm WD} }{0.15} \right) ^ {-2.63} \ \ \ {\rm d} \label{TREC}
\end{equation}

\noindent
where $M_{\rm donor}$ is the donor mass and 
$M_{\rm WD}$ is the WD mass. Notice that the smaller the mass ratio,
the longer the recurrence time.

The duration of the outburst is the interval between the beginning
and the end of the outburst and we adopt here the observational fit from
\citet{Smak_1999}

\begin{equation}
t_{\rm dur} \ = \ 2.01 \; \left( P_{\rm orb} \right)^{0.78} \ \ \ {\rm d} \label{TDUR}
\end{equation}

\noindent
where $P_{\rm orb}$ is the orbital period. Note that the longer
the orbital period, the longer the duration of the outburst.

\subsubsection{CVs with stable discs}
\label{stable_disc}

Since not all CVs are DNe, the CATUABA code also has to 
be able to deal with systems that have stable discs.
Now, we describe what happens in these cases.
In stable (hot or cold) discs, it happens that
the mass transfer rate is identical to the accretion rate.
We adopt the maximum values in both cases, again
using the prescription given by \citet{Lasota_2001} and 
\citet{Warner_1995_OK}, i.e.
for stable and cold discs, we use $\dot{M}_{\rm A}$,
and for stable and hot discs, we use $\dot{M}(d)$. Then,
the same procedure is executed for computing the luminosities
and the bolometric correction. Obviously, computations for the recurrence time and
duration of the outburst are not necessary.

\subsection{Probability of Detecting a CV}
\label{proba}

One of the objectives of this work is to compute the probability
of detecting a DN when observing a given GC at a distance 
$R_{\rm GC}$ associated with a given optical limiting 
magnitude $m_{\rm lim}$.

For that, let us first consider two principal configurations: 
(i) the CV can be detected by an optical detector/telescope in quiescence;
(ii) the CV can be detected by an optical detector/telescope in outburst;

Accordingly, we can have combinations of these two cases. 
Thereby, the chances of observing a given CV
will change based on such a combination.
Given a distance $R_{\rm GC}$ and a limiting magnitude $m_{\rm lim}$,
the probability ($P_{\rm obs}$) can simply be taken from the following. 
If the source can be detected only during outburst, then
$P_{\rm obs} = P_{\rm O} (t_{\rm dur},t_{\rm rec}) = t_{\rm dur}/t_{\rm rec}$ 
is the probability of detecting the outburst in one night\footnote{ 
The quantity $t_{\rm dur}/t_{\rm rec}$ corresponds to the very DN duty cycle.}.
If the source can be detected during quiescence and during outburst,
then $P_{\rm obs} = P_{\rm ecl} + P_{\rm O}(t_{\rm dur},t_{\rm rec}) = 1/3 + t_{\rm dur}/t_{\rm rec}$ 
is the probability of detecting such a CV, during the eclipse or during outburst, 
in one perfect night, where $P_{\rm ecl}$ is the probability of a CV being 
an eclipsing one. Finally, if the CV cannot be detected even 
during outburst, then the probability is null.

Notice that we are assuming that the only way to detect a CV
during quiescence is when the stars eclipse each other.
That is the why we add the $1/3$ term in the 
expression for the probability which corresponds to the 
probability of a binary being an eclipsing one 
\citep[e.g.][see appendix A2 for a demonstration]{Hurley_2002}.

In summary, we have the following law for the probability of detecting a DN
during its cycle:

\begin{equation}
    P_{\rm obs} \ = \
  \begin{dcases}
    t_{\rm dur}/t_{\rm rec},      & \text{if the CV can be} \\ 
                                  & \text{observed only during} \\
                                  & \text{outburst} \\
				  & \\
    1/3 + t_{\rm dur}/t_{\rm rec},& \text{if the CV can be} \\ 
                                  & \text{observed during} \\
                                  & \text{quiescence} \\
				  & \\
    0,                            & \text{otherwise}
  \end{dcases}
\label{PROBA}
\end{equation}

It is worth explaining some issues associated with the probability
expressed in equation \ref{PROBA}. We are not considering many of the
observational difficulties associated with observing a GC, like its intrinsic 
characteristic of being in a crowded field, a reduced number of
(good) nights dedicated to observing the GC thereof, instrumental
problems, etc.

Since the probability is computed mainly with the aim of dealing
with the apparent absence of DNe in GC, we do not have any prescription
for $P_{\rm obs}$ associated with CVs having stable discs.
Thus, equation \ref{PROBA} is associated with DNe only.

\subsection{Additional Considerations}
\label{facilities}

\begin{figure*}
   \begin{center}
    \includegraphics[width=0.85\linewidth]{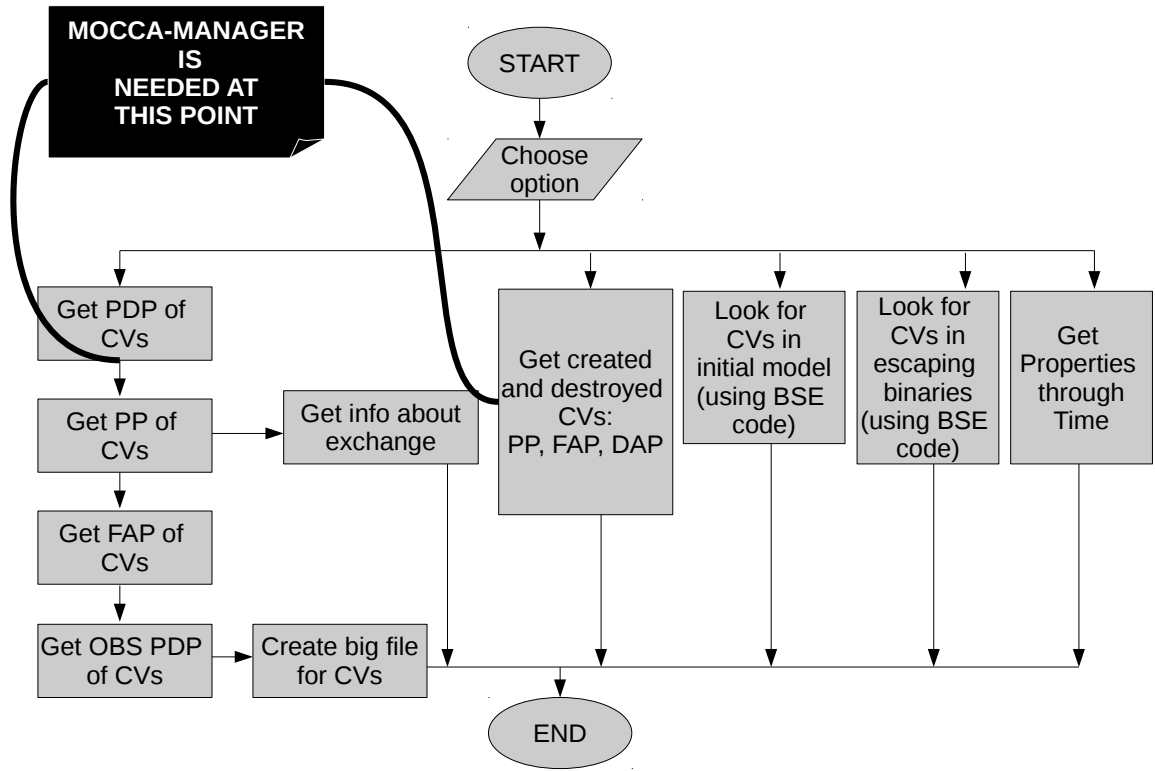} 
    \end{center}
  \caption{Flow chart of the CATUABA code. Notice that to get the info about progenitor population and formation-age population,
for CVs present (or not) in the cluster at 12 Gyr, the files containing
the very detailed history of the stars lives are needed. This is obtained by using
the MOCCA-MANAGER code. Notice that this flow chart reflects the underlying coding 
structure and is not meant to represent a logical flow or progression in the data 
analysis.}
  \label{Fig00}
\end{figure*}

So far, we have described the main aspects of the CATUABA code used
to study the PDP of CVs present in GCs simulated by MOCCA. Nonetheless,
other options are available in the code that allow for studying aspects
of the CVs not only included in the PDP. Now, we turn to such
options.

\subsubsection{Destroyed CVs}
\label{destroyed}

Throughout the evolution of a GC in the MOCCA code, not only are the CVs in the 
PDP formed, but many other CVs are formed as well. The only difference
is that they do not survive up to the present day. One of the capabilities of the
CATUABA code is the ability to collect relevant information about such systems,
in order to investigate their properties and the reasons for their
absence in the PDP.

This option gives similar information as given by the study of the PDP.
The main difference is that the information regarding the PDP is replaced
by the information associated with the destruction-age population which
correspond to the CV properties immediately before their disappearance from
the cluster.

With such a sample, one can investigate the properties of such 
`destroyed'\footnote{The reader should keep in mind that
the term `destroyed' used here is not necessarily associated
with a destructive process {\it stricto sensu}. The term
adopted here has a more general sense: CVs that are created
during the cluster evolution but do not survive up to the
present-day due to many reasons.}
CVs and also delineate general destruction channels.

In the possession of such information, we can separate the CVs into groups,
according to the main channels of `destruction' (similarly to what
is done for the CVs in the PDP).
Our choice on the matter is as follows: (i) 
{\it destruction due to binary stellar evolution group}: 
the CV stops being a CV due to some evolutionary process and
{\it without} any influence from dynamical interactions.
(ii) {\it destruction due to escape}: 
the CV life in the cluster is interrupted because the CV
escaped the cluster while still being (or not) a CV after the escape.
(iii) {\it destruction due to dynamical interaction group}: 
the death of the CV was caused by a dynamical
interaction, but the outcome of this interaction remains in the
cluster.

Notice that a CV can escape from the system due to either relaxation
processes or dynamical interactions. Both cases here are related to the
destruction due to escape group. In the case associated with CVs destroyed by dynamical
interactions without escaping, such CVs belong to the 
destruction due to dynamical interaction group. 

\subsubsection{Field-like CVs}
\label{field_like}

Two other important tools in the code are the possibility to
run the BSE code alone for the initial binaries and for the escaping binaries.
The former allows one to study the field-like population (without dynamics)
of CVs that comes from the initial binaries. This permits one to investigate how different
the cluster and field-like populations are, or even constrain the initial distributions 
of binaries
with available data from observed CVs in the solar neighbourhood. The latter
allows the study of the escaping CV evolution and the evolution of escaping binaries that
will become CVs up to 12 Gyr.

Again, for the BSE runs, similar information is recorded about: progenitor population, formation-age population,
PDP and destruction-age population. In the case when BSE evolves the initial binaries, information about the progenitor population,
formation-age population and PDP are saved for the CVs that survive up to 12 Gyr. If the CV
is somehow `destroyed', then the information about the destruction-age population is saved instead
of the information about the PDP. When BSE is run for the escaping
binaries, additional information is saved for the 
binaries immediately after the escape. This would correspond to the
escaping-age population. Again, information about the formation-age population is saved
always and, about destruction-age population or PDP, whether or not the CV survives up to 12 Gyr.

\subsubsection{Time-dependence of CV properties}
\label{time_dependence}

The last resource provided by the CATUABA code in its current version
is the possibility to investigate the time dependent properties of the CVs
from the cluster birth up to the present-day. Such an option in the code
utilizes the cluster snapshots. For a particular snapshot, all CVs
present in it are collected and their semi-major axis, period, mass ratio
 and mass distributions are generated. 
This is done for all the snapshots of the cluster (every 200-250 Myr, depending on the cluster).
It provides an efficient way to investigate how such distributions evolve with
time. Apart from the distributions,
the code also saves the average and median values of the semi-major axis,
period, mass ratio, and masses through time, again, based on all snapshots.

This option is available for the MOCCA snapshots and also for the BSE snapshots,
when one is interested in comparing both populations of CVs. In this case,
the timesteps for generating the snapshots of the cluster in MOCCA and BSE
have to be the same. In the case when only BSE snapshots
are being analysed, the timestep for the creation of such snapshots
can be any value. The timestep associated with MOCCA snapshots is defined before
the simulation and this value has to be used in the CATUABA code.

Here, we summarize all the possible ways of using the CATUABA code as
described above.

\subsection{Summary of CATUABA features}
\label{summary}

In order to account for the limitations of the above-presented approach,
we present a summary it and address the principal hypothesis which
should be kept in mind when interpreting the
results presented in this series of papers 
(especially, in Section \ref{probability} of this paper).
The study of the CVs is carried out in the following way.

(i) The first step is to recognize the PDP of the CVs in the snapshot
of the cluster at 12 Gyr. The next is to compile the most relevant
events in the history of all the stars in the PDP. With this,
we can unequivocally separate the CVs into three groups, namely:
BSE group, WDI group and SDI group.

(ii) The second step is to cover the shortcomings for CVs with 
stable or unstable discs. This is done by considering the 
formalism in the DIM. Such {\it modus operandi} provides good 
agreement with observation, and also allows for filling the deficiencies
left by BSE.

(iii) The third step is to compute two important time-scales of DNe, namely:
recurrence time and duration of the outburst. This is done by using
empirical relations.

(iv) The forth step is to calculate the probability of observing
and detecting the DNe by means of equation \ref{PROBA}.

It is rather important to mention at this point that we do not have any 
prescription for magnetic CVs. They correspond to roughly 25 per cent of CVs 
in the field \citep[e.g.][]{Wickramasinghe_2000,Ferrario_2015}
 and can be extremely important 
in GCs \citep[e.g.][]{Grindlay_1995,Dobrotka_2006,Ivanova_2006}. 

Secondly, most short-period DNe are SU UMa stars which also
have superoutbursts in addition to normal outbursts. 
Superoutbursts are more luminous (increase
in brightness by $\sim$ 1 mag), last longer (10 -- 20 days),
and are more infrequent (every 3 -- 10 DN cycles). 
In this way, the chances to observe a short-period DN during
superoutburst are even smaller since for one superoutburst,
we have 3 -- 10 outbursts.
Since we are interested here in the upper limit
of the probability, we neglect this complication.

Thirdly, the values for the accretion rate and, in turn,
luminosities and magnitudes, are derived based on maximum
values provided by the DIM. Besides, the mass transfer rate computed 
is a rough estimate since its based on average value of the donor 
mass loss during the CV evolution.

Finally, all information obtained in the CATUABA code is associated
with the BSE code. Thus, it is limited by what the BSE code is capable of.

To sum up, the principal hypothesis developed in this work includes the
neglect of magnetic CVs and superoutbursts for the short-period
DNe; the extrapolation of empirical relations for the CVs;
the assumption of theoretical maximum values for the
accretion rate, and consecutively, maximum luminosities.
All these assumptions have the aim of computing the 
upper limit for observing the CVs.

It is worth noticing that the procedure described above is somewhat forged,
although reasonable. This is because we cannot retrieve precisely
the CV properties from observations.

Other options can be used independently from 
what is done for the CVs in the PDP (Section \ref{facilities}). A flow chart
of the code is presented in Fig. \ref{Fig00}. In the very beginning,
the user has to choose an option which includes

(i) getting the PDP, the progenitor population,
the formation-age population and the observational properties for the CVs present in the cluster
at 12 Gyr; 

(ii) getting the progenitor population, escaping-age population, formation-age population and destruction-age population for the
CVs formed during the cluster evolution but not present in the PDP;

(iii) getting the progenitor population, formation-age population and destruction-age population/PDP for the field-like CVs; 

(iv) getting the progenitor population, formation-age population and destruction-age population/PDP for the CVs created from escaping binaries; 

(v) getting the distributions of the main orbital elements of the CVs though time.

Notice that this flow chart 
reflects the underlying coding structure and is not meant to represent 
a logical flow or progression in the data analysis.

After introducing the CATUABA code and also the hypothesis in it, 
we are able to turn to the description of the six models (Section \ref{models}) 
that are used in these initial investigations and 
the main results on the PDP of CVs in such models (Section \ref{results}).

\section{MODELS}
\label{models}

\begin{table*} 
\centering
\caption{Models and parameters that define them.
The name of each model has a letter and a number.
The letter indicates its initial binary population
 as well as its initial binary fraction,
namely: K (Kroupa) with high initial binary fraction
and S (Standard) with low initial binary fraction
(see the text); and the number indicates its central density:
1 (sparse), 2 (dense) and 3 (very dense).}
\label{Tab01}
\begin{adjustbox}{max width=\textwidth}
\noindent
\begin{threeparttable}
\noindent
\begin{tabular}{l|c|c|c|c|c|c|c|c|c|c|c|c|c|c}
\hline\hline
Model & \specialcell{Mass\\$\left[ {\rm M}_\odot \right]$} & \specialcell{Number of\\objects}   &  \specialcell{Initial\\binary fraction} & \specialcell{Central Density\\$\left[ {\rm M}_\odot \; {\rm pc}^{-3} \right]$} &  \specialcell{$r_{\rm t}$\\$[{\rm pc}]$}  & \specialcell{$r_{\rm h}$\\$[{\rm pc}]$} & Z & IMF \tnote{a} & q \tnote{b,c} & a \tnote{b,c} & e \tnote{b,c} & \specialcell{Present-day\\Type \tnote{d}}  & \specialcell{Present-day\\binary fraction} \\ 
\hline\hline
K1 & \hspace{0.2cm} $8.07 \times 10^5$ & \hspace{0.2cm} $1.12 \times 10^6$ &  \hspace{0.2cm} $95$ \% & \hspace{0.2cm} $1.9 \times 10^2$ & \hspace{0.2cm} $115$ & \hspace{0.2cm} $16.9$ & $0.001$ & $3$ & Kroupa & Kroupa & Kroupa & pc & $28.3$ \% \\ \hline
K2 & \hspace{0.2cm} $8.07 \times 10^5$ & \hspace{0.2cm} $1.12 \times 10^6$ &  \hspace{0.2cm} $95$ \% & \hspace{0.2cm} $7.8 \times 10^4$ & \hspace{0.2cm} $115$ & \hspace{0.2cm} $2.3$ & $0.001$ & $3$ & Kroupa & Kroupa & Kroupa & cIMBH & $8.9$ \% \\ \hline

K3 & \hspace{0.2cm} $8.07 \times 10^5$ & \hspace{0.2cm} $1.12 \times 10^6$ &  \hspace{0.2cm} $95$ \% & \hspace{0.2cm} $3.2 \times 10^5$ &  \hspace{0.2cm} $72$ & \hspace{0.2cm} $1.4$ & $0.001$ & $3$ & Kroupa & Kroupa & Kroupa & cIMBH & $5.4$ \% \\ \hline
S1 & \hspace{0.2cm} $5.92 \times 10^5$ & \hspace{0.2cm} $1.00 \times 10^6$ &  \hspace{0.2cm} $05$ \% & \hspace{0.2cm} $2.8 \times 10^3$ &  \hspace{0.2cm} $97$ &  \hspace{0.2cm} $7.5$ & $0.00016$ & $3$ & uniform & log-normal & thermal & c & $4.9$ \% \\ \hline
S2 & \hspace{0.2cm} $9.15 \times 10^5$ & \hspace{0.2cm} $1.80 \times 10^6$ &  \hspace{0.2cm} $10$ \% & \hspace{0.2cm} $1.3 \times 10^5$ & \hspace{0.2cm} $125$ &  \hspace{0.2cm} $2.1$ & $0.001$ & $2$ & uniform & \specialcell{uniform\\in $\log(a)$}  &  thermal & pc & $4.8$ \% \\ \hline
S3 & \hspace{0.2cm} $1.52 \times 10^5$ & \hspace{0.2cm} $3.00 \times 10^5$ &  \hspace{0.2cm} $10$ \% & \hspace{0.2cm} $5.9 \times 10^5$ &  \hspace{0.2cm} $69$ &  \hspace{0.2cm} $0.7$ & $0.001$ & $2$ & uniform & \specialcell{uniform\\in $\log(a)$} & thermal & c & $4.6$ \% \\ \hline \hline
\end {tabular}
\begin{tablenotes}
       \item[a] The IMF 3 is the Kroupa IMF with three segments \citep{Kroupa_1993} and the IMF 2 is the Kroupa IMF with two segments \citep{Kroupa_1991}.
       \item[b] The Kroupa IBP corresponds to the construction of the parameters based on the eigenevolution and the mass feeding algorithm \citep{Kroupa_INITIAL}.
       \item[c] The Standard IBP is associated with uniform distribution for the mass ratio, uniform distribution in log or log-normal one for the semi-major axis and the eccentricity thermal distribution is such that it follows from a uniform binding-energy distribution.
       \item[d] The cluster present-day type can be: post-core collapse (c), post-core collapse with intermediate-mass black hole (cIMBH) and pre-core collapse (pc).
\end{tablenotes}
\end{threeparttable}
\end{adjustbox}
\end{table*}

In this first examination of CVs in GCs, we have chosen six models that
differ mainly with respect to the initial central density, initial binary distributions,
and initial binary fraction. The restricted number of models used in this
work is due to its very objective nature: to construct the numerical apparatus used 
to study CVs based on MOCCA simulations, to check its consistencies, and to
gain experience for data analysis.

Before proceeding further, it is convenient to define the {\it initial binary
population} (IBP) since it is a rather important concept
in this work. The initial binaries in GCs follow determined
distributions for their parameters: semi-major axis,
eccentricity, masses, mass ratio, and period.
Hereafter, the initial binaries associated with the initial
distributions for their parameters belong to the IBP. 
In other words, the IBP
is the group that contains all initial binaries, in a given initial
cluster, associated with specific initial distributions for
their parameters. In this work, we analyse models with two distinct
IBPs.

The first set of models, defined as the K models,
corresponds to models constructed based on the IBP 
derived by \citet{Kroupa_1995,Kroupa_INITIAL,Kroupa_2013}.
Such an IBP has the following properties:
(i) the period distribution favours long-period binaries; (ii)
the orbits tend to be circularized for short-period binaries,
and (iii) the mass ratio distribution
is almost flat with a huge peak close to $\sim 1.0$ \citep{Marks_2011}.
The K models have 95 per cent primordial binaries\footnote{
We set the initial binary fraction for the models with Kroupa IBP
different from 100 per cent in order to avoid computational problems 
that arise in MOCCA if there is no single star in the initial
model.}, the same initial mass $8.07 \times 10^5 \ {\rm M_\odot}$ and 
the same initial number of objects (binaries and single stars)
$1.12 \times 10^6$. But they differ with respect to the initial central
density, having, in ${\rm M}_\odot \; {\rm pc}^{-3}$, $1.9 \times 10^2$,
$7.8 \times 10^4$ and $3.2 \times 10^5$. Thus, we have
one sparse model, one dense model and one very dense model.

The other set of models, defined as S models, follows the `Standard'
IBP and has low binary fractions of 5 and 10 per cent.
The Standard IBP is associated with
a uniform distribution for the mass ratio, a uniform log or log-normal
distribution for the semi-major axis, and a thermal distribution for the
eccentricity.  As in the K models, we have chosen models with different
initial central densities, namely: $2.8 \times 10^3$, $1.3 \times 10^5$ and
$5.9 \times 10^5$, in  ${\rm M}_\odot \; {\rm pc}^{-3}$.
Again, for the set of S models, we have a sparse model, a dense model and 
a very dense model. Differently from the K models, in the S models,
we have different initial numbers of objects and initial masses.

We have used two IMFs that follow the 
broken power law $\xi(m) \propto m^{-\alpha}$, defined by 
\citet{Kroupa_1991,Kroupa_1993}. The IMF2 (canonical) is such that it has
$\alpha = 1.3$ for $0.08 \leq m/{\rm M_\odot} \leq 0.5$ and 
$\alpha = 2.3$ for $0.5 \leq m/{\rm M_\odot} \leq m_{\rm max}/{\rm M_\odot}$ \citep{Kroupa_INITIAL}. 
The IMF3 (multiple power-law) is such that it has 
$\alpha = 1.3$ for  $0.08 \leq m/{\rm M_\odot} \leq 0.5$, 
$\alpha = 2.3$ for $0.5 \leq m/{\rm M_\odot} \leq 1.0$ and
$\alpha = 2.7$ for $1.0 \leq m/{\rm M_\odot} \leq m_{\rm max}/{\rm M_\odot}$
 \citep{Kroupa_INITIAL}.
The star mass in this study lies between $0.08 {\rm M_\odot}$ and $100 {\rm M_\odot}$. 

We assume that all stars are on the zero-age main sequence
when the simulation begins and that any residual gas from the star
formation process has already been removed from the cluster. Additionally,
all models have low metallicity, are initially at virial equilibrium, and have neither
rotation nor mass segregation. Moreover, all models are evolved 
for 12 Gyr which is associated with the present-day in this investigation.

In Table \ref{Tab01} we summarize the main parameters of the initial 
models. 

In addition to the evolution done using MOCCA, we have also evolved 
the initial models (actually, the IBP) with
BSE alone during 12 Gyr. In such a way, we will have information not only about the
cluster CVs but also on the field-like CVs. This will help us to compare
both groups of CVs that come from the six initial models.

\section{RESULTS AND DISCUSSION}
\label{results}

In this section, we present the main results related to the 
PDP of CVs in the six models described in Section \ref{models}.
First, we will discuss some evolutionary properties of the 
models. Secondly, we will state the properties of the PDP
of CVs. After that, we will show how observational
selection effects can hide the CV population in observed
GCs by computing the upper limit of observing such CVs. 
We also address some discussions on the main
points and correlate this work with previous ones.
Finally, we discuss possible observational procedures
that might help in detecting the missing DN population.

\subsection{Cluster Evolution}
\label{cluster}

The clusters were evolved to an age of roughly 12 Gyr -- considered here 
as the present-day -- which required one day in the PSK cluster
at the Nicolaus Copernicus Astronomical Centre (CAMK) in Poland.

The last two columns of Table \ref{Tab01} show the properties of the clusters
at the present-day. Notice that for six models, 
we have relatively representative models, including sparse, dense and 
very dense clusters; high and small initial binary fraction;
post core collapse clusters and two with intermediate-mass
black holes.

Also important is that the binary fraction in models K1, 
K2 and K3 drops drastically in the early evolution of the clusters,
since the soft binaries tend to be disrupted.
This is due to the interplay between the cluster densities and
the initial distribution of the period in the Kroupa IBP which
is dominated by soft binaries.
Besides, the denser the cluster, the faster the decrease
in the number of binaries (see the last column of Table \ref{Tab01}). 

Previous studies on CV and also on AM CVn (interacting binary in which a WD accretes
matter from another WD) were performed on the basis of the Standard IBP 
(see Section \ref{models}). 
This is the first work which includes not only such Standard
IBP but also the Kroupa IBP. Interesting results come from the 
Kroupa IBP \citep[e.g.][]{Kroupa_1995,Kroupa_INITIAL} which can reproduce,
to some extent, some observable properties of the binaries in the Galactic field.
Nevertheless, the Kroupa IBP was never tested with respect to the observed
CV properties.
In this way, analysing such models will allow for the first time checking
the consistency of the Kroupa IBP regarding CVs, especially those in the
field since most stars in the field should have come from the dissolution
of star clusters \citep[e.g.][]{Lada_2003}.

\subsection{Dependence on Initial Model}
\label{cv_model_dependence}

We can start the analysis of the PDP of CVs by providing the number
of CVs in each model separated by the three groups described in 
Subsection \ref{categorisation}. This is given in Table \ref{Tab02}.
We also included in the table the CVs formed due to exchange. In addition, 
we include the number of binaries in the progenitor population that become CVs when BSE is run alone
, i.e. binaries that would become CVs no matter the number/strength of dynamical
interactions, if any took place. Finally, in the last column of the 
table we state the number of CVs formed when evolving the IBP
 out of the cluster evolution, i.e. evolving all the initial binaries
with the BSE code alone. This is the field-like population of CVs
 that comes from the respective IBP.

\begin{table}
\centering
\caption{Number of CVs that are present in the clusters at 12 Gyr
separated by the group to which they belong, namely: binary stellar evolution (BSE) group, 
weak dynamical interaction (WDI) group or strong dynamical interaction (SDI) group. 
Also indicated is the number of CVs for which exchange is the main
channel for their formation as well as the number of progenitors of the CVs
(binaries in the progenitor population) that 
become CV when BSE code is run alone, i.e. that become CVs purely via 
binary evolution without the cluster environment. The last column indicates the number
 of CVs obtained when BSE evolves all the initial binaries and correspond to the number
of field-like CVs that come from the respective IBP.}
\label{Tab02}
\begin{adjustbox}{max width=430px}
\noindent
\begin{threeparttable}
\noindent
\begin{tabular}{l|c|c|c|c|c|c|c}
\hline
Model & BSE \tnote{a} & WDI \tnote{b} & SDI \tnote{c} & Total & Exchange & CVF \tnote{d}  & Field-like  \tnote{e}\\ 
\hline\hline
K1 &   0 &   0 &   3 &   3 &    2 &   0  &   0 \\ \hline
K2 &   0 &  18 & 164 & 182 &  120 &   0  &   0 \\ \hline
K3 &   0 &   9 & 121 &  130 &  98 &   0  &   0 \\ \hline
S1 &  40 &   1 &   1 &   42 &   1 &  41  &  69 \\ \hline
S2 & 116 & 113 &  31 &  260 &  13 & 197  & 239 \\ \hline
S3 &   9 &  14 &   4 &  27 &    2 &  21  &  64 \\ \hline 
\hline 
\end{tabular}
\begin{tablenotes}
       \item[a] CVs that are formed due to Binary Stellar Evolution only.
       \item[b] CVs that are formed with the influence of Weak Dynamical Interaction.
       \item[c] CVs that are formed with the influence of Strong Dynamical Interaction.
       \item[d] CVs formed when BSE code evolves the progenitors of the CVs (binaries in \\
		the progenitor population).
       \item[e] CVs formed when BSE code evolves all the initial binaries.
\end{tablenotes}
\end{threeparttable}
\end{adjustbox}
\end{table}

\begin{figure*}
   \begin{center}
        \includegraphics[width=0.45\linewidth]{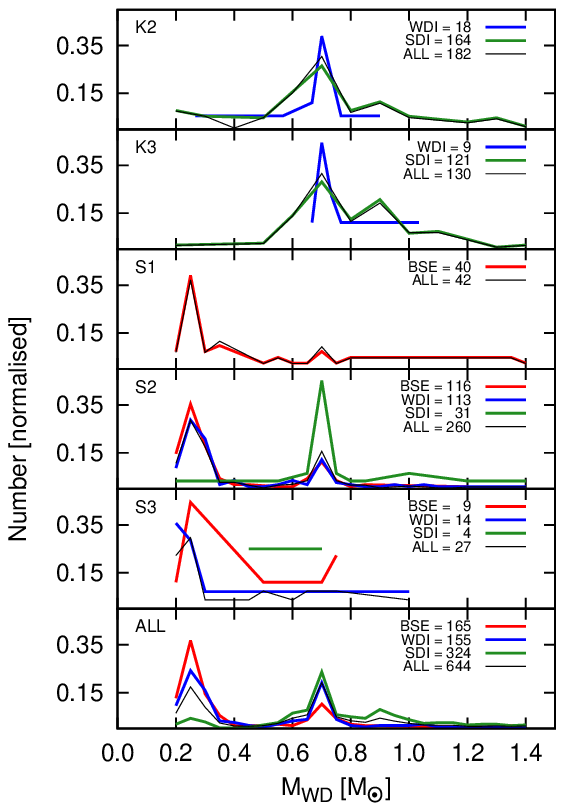} 
        \includegraphics[width=0.45\linewidth]{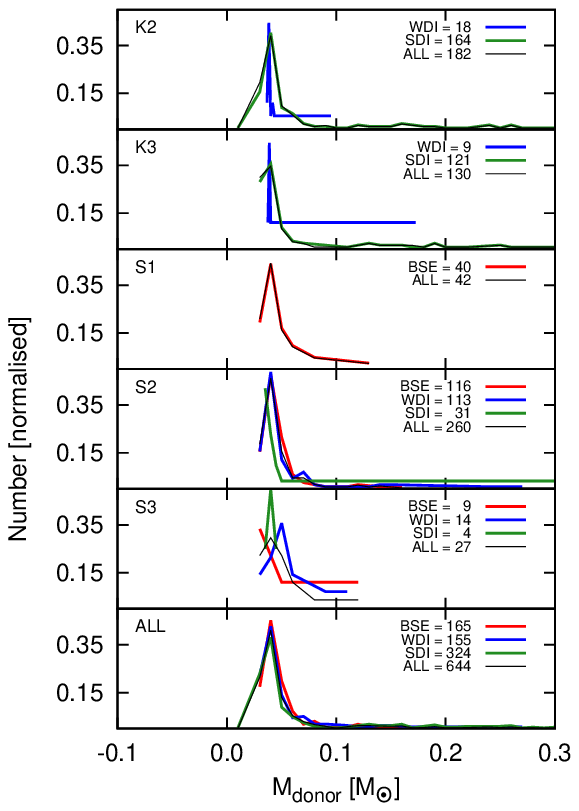}
    \end{center}
  \caption{ Present-day CV component masses, normalized by the total number indicated
in the key. {\it Left}: the WD mass with distinction for the 
groups of CVs, namely BSE, WDI and SDI.
Notice that the CVs in the SDI group have masses peaking around
$0.7 {\rm M_\odot}$ and the BSE group contains CVs whose WD mass is peaked
around $0.3 {\rm M_\odot}$.
{\it Right}: the donor mass. Notice that for the donor, we do not
observe any difference associated with the influence of dynamics, i.e.
the three groups have very low-mass donors, in general. 
The keys indicate the number of CVs in each model and group and the
last panel represents the PDP of CVs present in all six models
aggregated as one.}
  \label{Fig03}
\end{figure*}

\subsubsection{Absence of field-like CVs in Kroupa models}
\label{KIBP}

It is clear from the table that there is no CV produced via only 
binary stellar evolution in the model with Kroupa IBP.
In other words, in order to form CVs in the models with Kroupa IBP 
(K1, K2 and K3),
dynamical interactions have to take place. This is a strong hint
towards an inconsistency between observations and theoretical predictions,
because we do observe CVs in the field and regions where dynamics could 
not have played a role. Either we are missing important physics, or the 
assumption of a Kroupa IBP is not the right one here.

It is thought that the field population of stars comes from the dissolution
of star clusters \citep[e.g.][]{Lada_2003,Marks_2012} after the expulsion
of the residual gas in the star formation process.
The simulations here were performed such that the initial clusters are set
after the removal of such residual gas. Thus, if the cluster is 
quickly dissolved due to mass removal
because of super novae and stellar evolution of the most massive stars, 
the binary parameters would not 
be strongly modified (particularly the short-period ones that could be
progenitors of CVs).
So such a population that comes from the dissolved clusters should represent 
the field population. In other words, dissolved clusters associated with 
Kroupa IBP should correspond to the field parent population of binaries
\citep{Kroupa_1995}.
Hence, since they have around $10^6$ initial binaries,
we should expect a reasonable number of CVs formed
from them due to stellar binary evolution only.

Surprisingly, not a single CV was produced in models K1, K2 and K3; even when 
evolving the IBP with BSE alone. The reason could be: either
the mass feeding algorithm that increases the mass ratio of the binaries that would
be potential progenitors of the CVs might require adjustments, or the
efficiency of the CEP should be much smaller.

With respect to the mass feeding, the mass ratio associated with the progenitor
binaries of CVs is small (q $\lesssim$ 0.2) and they
are short-period binaries. Since 
the mass feeding procedure tends to increase the secondary
mass due to accretion of gas from the circumbinary 
material, the secondary mass increases while the primary mass 
remains constant \citep{Kroupa_INITIAL}. This implies 
that the initial mass ratio of short-period binaries increases 
towards 1. In this way, it seems that the feeding procedure generates 
initial binaries that are inappropriate for evolving into CVs, 
independently of their orbital period distributions.

On the other hand, this difficulty would be overcome 
if the CEP efficiency decreases significantly.
Then a great deal of orbital energy would be needed in order
to eject the common envelope; this would bring the long-period
binaries (with low q) to shorter periods, and consecutively 
turn them into potential CVs. Currently, the adopted value
is $\alpha = $ 3.0. By reducing this value to $\lesssim$ 1.0, 
some CVs could be formed.

The Kroupa IBP has been checked against numerical simulation and observations,
and has successfully explained observational features of young clusters,
associations, the Galactic field, and even binaries in old GCs
\citep[e.g.][and references therein]{Kroupa_2011,Marks_2012,Leigh_2015}. 
Nevertheless, our results show that the Kroupa IBP cannot form CVs without
the influence of dynamical interactions, given the adopted parameters
in BSE.

We will leave for a separate paper a more elaborate discussion on the Kroupa
IBP and the stellar binary parameters.

\subsubsection{Influence of dynamics}
\label{KIBP}

Yet with respect to Table \ref{Tab01}, we see from the last two
columns how dynamics are important regarding CVs in GCs. Since these
columns represent the number of CVs formed without the influence of
dynamics and survive up to the present-day, we can see somehow the
influence of dynamics in the S models.

For model S1, almost all CVs are formed from primordial binaries 
(compare the fifth and seventh columns). This is because this model is sparse. 
Additionally, we see that some CVs that are
formed in a field-like environment are missing in the cluster (compare the seventh and
last columns). This indicates that dynamics in such a cluster contribute to lower
the number of CVs produced in the cluster environment by destroying either
their progenitors or the CVs themselves.

For models S2 and S3, we see that not all CVs in the PDP come from primordial
binaries (compare the fifth and seventh columns). This indicates that 
some CVs were formed dynamically. Besides, some CVs or some
of their progenitors are destroyed
in the cluster environment (compare seventh and last columns) like
in model S1. Thus, dynamics in models S2 and S3 contribute to both
the creation and destruction of CVs.

For models K1,K2 and K3 we have formation only with
the influence of dynamical interactions. Specifically, exchange being the most important
channel for the formation of CVs in these clusters, which happen for $2/3$ of the
entire PDP of CVs. Then, for the sparse cluster K1, the `dynamical' process of formation is not
expressive and only three CVs could form in such a cluster. However, for the denser
clusters (K2 and K3) in the set, this is not true. They have a reasonable number of
CVs formed because of dynamical interactions since such interactions
took place more frequently in K2 and K3 than in K1.
 
Unfortunately, more details about how the initial density and initial binary 
populations influence the CV formation are not obtained by only analysing 
the PDP of CVs in each model. In fact, we need information about the
production/destruction of CVs through the cluster evolution as a whole,
i.e. the complete formation rate of CVs in the models.

\subsection{CV properties}
\label{cv_properties}

Now, with respect to the CV properties themselves,
in Fig. \ref{Fig03}, the WD and donor masses of the CVs 
at the present-day are displayed. We also included in the figure (last panel) the PDP
of all six models aggregated as one.

\subsubsection{WD mass}
\label{cv_wd_mass}

We can see clearly a distinction between CVs strongly influenced
by dynamics and CVs which were not strongly influenced. The former has a peak
around $0.7 {\rm M_\odot}$ while the last has a peak around $0.3 {\rm M_\odot}$
for the WD mass. Additionally,
the former belong to the SDI group while the latter to the BSE group. In this way,
dynamics bring more massive accretors to the PDP. This is because of the exchanges
that take place in the CVs belonging to the SDI group.
In this case, it is more probable to exchange a low-mass star 
for a more massive star in dynamical interactions.
Thus, the future accretor will be more massive.

Concerning the WDI group, if
the cumulative effect of the weak interactions in the binary that becomes CV is strong, 
then, such a CV will have properties similar to those in the SDI group, as in the 
case of the CVs in models K2 and K3. On the other hand, if the net effect of the weak
interactions is soft, then the CV will have similar properties as those of
the BSE group. This can be seen in models S2 and S3.

Generally, with respect to all analysed models,
each CV in the WDI group suffered few weak dynamical interactions
before the CV formation. Around 37 per cent of the CVs in the WDI group
suffered only one dynamical interaction,
$\sim$ 20 per cent suffered two and $\sim$ 10 per cent suffered three weak interactions. The rest
of them suffered more than three weak interactions. 
Surprisingly, there is one CV that went through 28 weak dynamical interactions.
Now, after the CV formation,
most of them suffered only one weak dynamical interaction which changes slightly
the eccentricity by an amount of $\sim 10^{-8}$.

Concerning the CVs (in all six models) in the SDI group, they suffered
few strong dynamical interactions before CV formation.
Most of them ($\sim$ 71 per cent)
experienced only one strong dynamical interaction. Around 20 per cent
underwent two strong dynamical interactions and $\sim$ 6 per cent
of them went through three strong dynamical interactions. The rest of them
($\sim$ 3 per cent) suffered four or more strong dynamical interactions.
There is one CV that went through seven strong dynamical interactions
before CV formation.
Regarding the evolution after CV formation, only one CV in one model
went through one strong dynamical interaction.

\subsubsection{Donor mass, period and CV status}
\label{cv_donor_mass_period}

In Fig. \ref{Fig03}, we see no difference among these groups with
respect to the donor mass. The majority of the CVs in the six models have
BD companions. This means that they are mainly in the last stage
of CV evolution ($\sim$ 87 per cent are period bouncers with BD as donor).
The reason for that
is simple, the CVs are formed mainly in the domain of short-period CVs which means
that they will easily evolve up to the period bouncers domain.

Obviously, not all CVs in the PDP are period bouncers. 
The youngest CVs are still short-period CVs, long-period CVs or CVs in the gap.
Nevertheless,
considering 12 Gyr of cluster evolution, it is reasonable to have most of them
as period bouncers. This has already been shown by \citet[][]{Howell_1997},
for instance, for CVs in the Galaxy.

Apart from the masses of the stars in a CV, the period has come out 
to be one of the most important observables on the matter. For the PDP
of CVs in the six models, we do not observe the period-gap (most of the
 CVs are formed below or in the gap) and the period distributions peak
around 3 h (in the period bouncers domain).

At this point, we can state the main characteristic of the PDP
of CVs in GC: {\it they are in (or close to) the last stage of their evolution},
as one can see from Fig. \ref{Fig03}.
This not only defines their properties but also
indicates that observational selection effects can be rather strong and have to be taken
into account for a consequential understanding of CVs in GCs
\citep[e.g.][]{Pretorius_2007,Knigge_2012MMSAI}.

After a brief discussion on the CV properties, we can concentrate
on the observational selection effects associated with the PDP of CVs in the 
six models which, in turn, is totally bound to the PDP thereof.

\begin{figure}
   \begin{center}
    \includegraphics[width=220px]{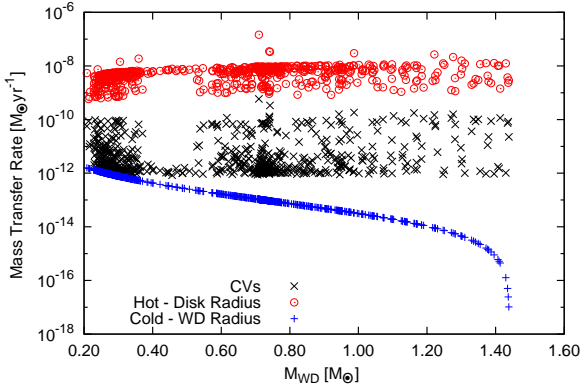}
    \end{center}
  \caption{Comparison of the values of the mass transfer rate required 
for a full CV disc to be globally hot and stable (circles) 
or globally cold and stable (plus signs) with
respect to the mass transfer rate of CVs in the PDP (crosses). 
One can notice that all CVs are DNe and the mass transfer rates of the 
CVs are such that it allows for the instability somewhere in the disc. It is
also possible to realize that the instability for many CVs takes place
very close to the inner edge (crosses close to the plus signs).}
  \label{Fig04}
\end{figure}

\begin{figure*}
   \begin{center}
    \includegraphics[width=0.45\linewidth]{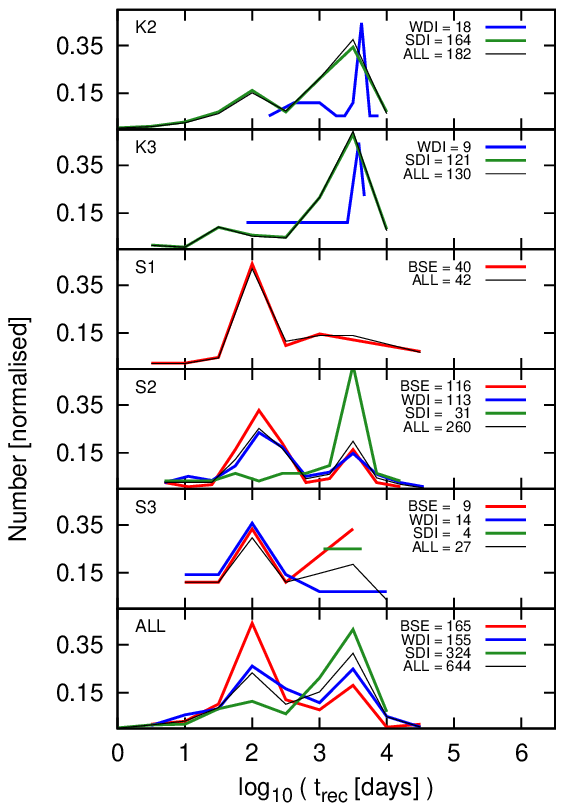}
    \includegraphics[width=0.45\linewidth]{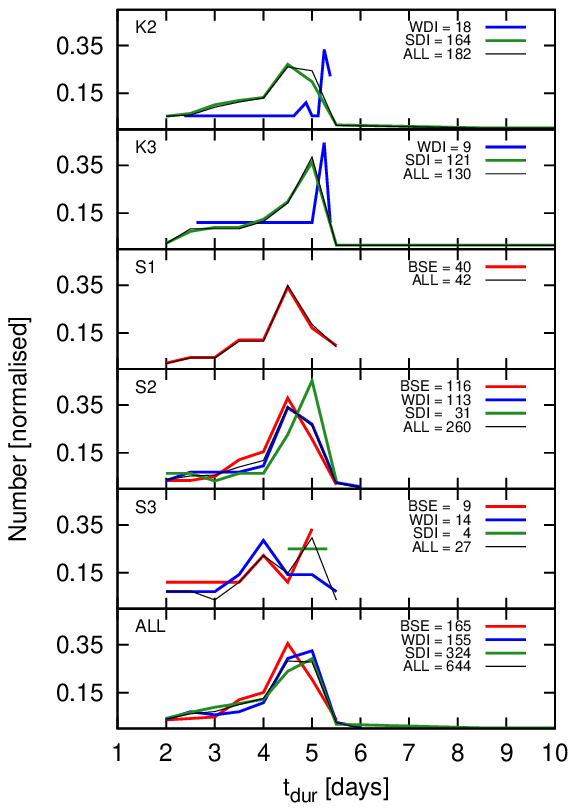}
    \end{center}
  \caption{Recurrence and duration times associated with the DNe
CVs present in the PDP. Notice that the outbursts last a few days and
the intervals between them are from a few days to decades. Notice a double
peak in the recurrence time distribution due to the double peak in
the WD mass distribution. Dynamically formed CVs have greater WD masses, then longer recurrence
times according to Equation \ref{TREC}. Now, for the duration of the outburst distribution, 
we do not see too great of differences between
dynamically formed CVs and CVs formed without influence of dynamics. This
is due to the CV status at the present-day (period bouncers). Most of them have 
periods from $\sim$ 2h to $\sim$ 3h, and according to Equation \ref{TDUR} (which is 
a function of the orbital period), they are similar.}
  \label{Fig05}
\end{figure*}

\subsection{Are CVs in GCs DNe?}
\label{cv_dne}

\begin{figure*}
   \begin{center}
    \includegraphics[width=0.45\linewidth]{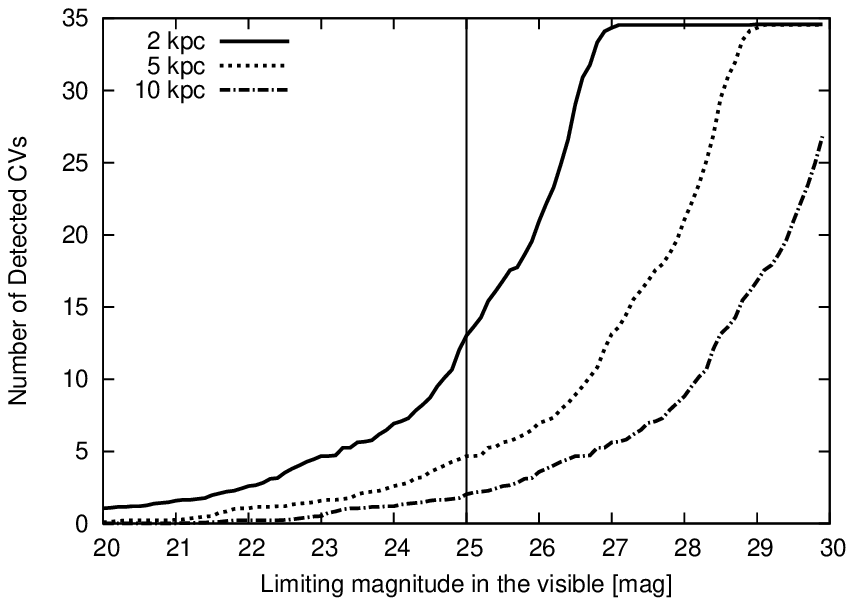}
    \includegraphics[width=0.45\linewidth]{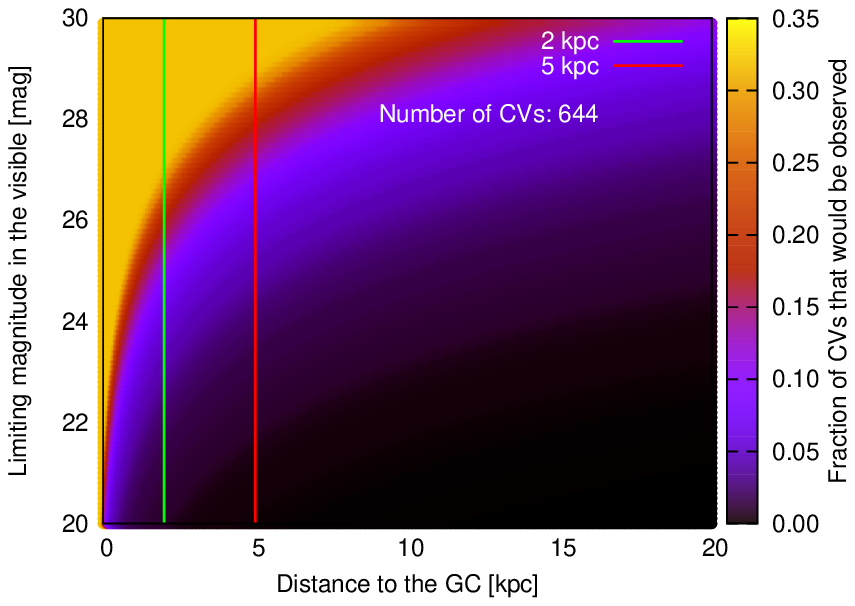}
    \end{center}
  \caption{{\it Left-hand panel}: number of CVs that could be observed in an average GC
as a function of the limiting magnitude for three different cluster distances.
{\it Right-hand panel}: CV detection rate as a function of the limiting 
magnitude and the cluster distance, considering all CVs in the 
six models. The colour bar represents 
the CV fraction that would be observed based on probability-weighted
2D normalized histograms in each cell of size 0.005$\times$0.1.}
  \label{Fig06}
\end{figure*}

One of the objectives of this work on CVs and related objects is to quantify the 
influence of observational selection effects on the analysis and prediction
done by models. In this sense, we decided to implement in the
CATUABA code an upper limit for the probability of detecting
CVs as a function of distance and limiting magnitude. This choice
was motivated by the aim of showing that, if we find small chances
to detect CVs, even in a perfect situation, then, in real situations,
it will be worse. This seems to be an extremely important
topic and should be associated with the so-called
absence of DNe in GCs.

The first step in the direction of computing the fraction of
CVs expected to be observed during an ideal night with an ideal
instrument is checking if the CVs in the PDP are DNe or not.
Using equations \ref{MA} and \ref{MB}, we can separate the CVs into three
ranges with respect to the disc stability as described in
Section \ref{cv_type}. Fig. \ref{Fig04} shows
that all CVs in the six models are potential DNe. Additionally,
the figure illustrates clearly that for a great deal of them,
the instability takes place quite close to the WD surface
(crosses close to field plus signs).
This indicates that the strength of the WD magnetic field does not
have to be very strong in order to prevent the disc instability and,
in turn, the outbursts in such systems. Fortunately, this is not
bad for our purposes since we are trying to derive an upper limit
for the real probability of detecting the CVs. In this way, having 
a disrupted disc (cold and stable) would lower the probability
even further.

Once we know that the CVs are DNe, we are able to compute the 
recurrence time and the duration of the outburst, based on equations
\ref{TREC} and \ref{TDUR}. The results are displayed in Fig. 
\ref{Fig05}.
Notice that the recurrence time varies from days to decades
and the outbursts last a few days. Notice also a double
peak in the distribution of the recurrence time due to the double peak in
the WD mass. Dynamically formed CVs have greater WD masses, implying longer recurrence
times according to Equation \ref{TREC}. Now, for the distribution
of the duration of the outburst, we do not see too many differences between
dynamically formed CVs and CVs formed without influence of dynamics. This
is due to the status of the CVs at the present-day (period bouncers). They mainly have  
periods from $\sim$ 2 to $\sim$ 3h and according to Equation \ref{TDUR} (which is 
a function of the orbital period) they are similar.

From Fig. \ref{Fig05},
we can see that the chances to observe a CV during outburst
are high only for some CVs -- the youngest ones that were formed
close to the present-day and have greater donor masses and 
higher mass transfer rates with respect to the old ones.
In this way, we have the first clue that a good solution would, instead
of looking for variability during an outburst, look for it during
quiescence. The biggest  problem associated with this approach is the need 
of very deep observations because in order to reach magnitudes
similar to those associated with a CV during quiescence,
one needs deeper observations in comparison with the same CV during 
its outbursts. Although a difficult task, it is possible.
For instance, \citet{Cohn_2010} reached
apparent visual magnitudes as deep as $\sim$ 26 when studying 
the optical counterparts of X-ray sources in the GC NGC 6397.

\subsection{Probability of Detection through Variability}
\label{probability}

Now, once we have separated the CVs with respect to the disc
instability and computed their recurrence times as well as
the duration of their outbursts, we can turn to the computation
of the probability of detecting such CVs by using equation \ref{PROBA}.

The left-hand panel of Fig. \ref{Fig06} displays the number of CVs that would
be observed as a function of the limiting magnitude for three different 
cluster distances, considering the PDPs in all six models (644 CVs). Thus, the number
of CVs indicated in the figure is related to the number of CVs that would be
observed in an `average' GC since it is associated with the total number of CVs seen
in all models divided by the number of clusters (i.e. six). In this average
cluster, the total number of CVs is $\sim$ 108.
For instance, for a limiting magnitude of 25 mag, at 2 kpc, one would observe
$\sim$ 13 CVs. For 5 and 10 kpc, one would observe $\sim$ 5 and $\sim$ 2, respectively.
As the limiting magnitude increases, the number of observed CVs increases up to 
a maximum of $\sim$ 1/3 (fraction of eclipsing CVs).

The right-hand panel of Fig. \ref{Fig06} is a generalization of the left-hand panel
and exhibits the CV fractions that would be
detected weighted by the probability of observing each CV in each 
cell of size 0.005$\times$0.1, computed based on normalized histograms 
in the plane of limiting magnitude versus distance, including all CVs
in the PDP of all six models. 
Notice that we have basically three regions in the figure: one dark,
one light and one in the middle. The dark region corresponds to CVs that
could be observed only during outburst since the probability is small.
The light region is associated with the CVs that can be observed also
during quiescence, because the probability is $\sim$ 1/3 (probability
of being eclipsing binary). Finally the middle region (not so dark nor
so light) corresponds to a mix of CVs that can be observed during quiescence
and during outburst.

In this way, the smaller the distance and the greater the limiting magnitude of 
the instrument, the greater the chances are to detect the CV during quiescence. 

Now we can compare this result with observations and check
its consistency. First, if one fixes the position of the cluster, let us say at
2 kpc (corresponding to the distance of the closest GCs and indicated in the figure),
and goes from lower limiting magnitudes
up to higher limiting magnitudes, we see that the probability is small
($\lesssim$ 10 per cent), up to about $\sim$ 25 mag. Then, from that point
up to $\sim$ 26.6 mag, the probability increases to $\sim$ 20 per cent.
From there, the probability reaches the maximum ($\sim$ 1/3).

As an example for comparison, 
\citet[][see their fig. 3]{Cohn_2010} reached a limiting magnitude of around 26 mag in the
R filter (similar filter to what we have) in their observations of NGC 6397.
In such a study, they found 15 candidates of CVs.
In this way, since the predicted number of CVs in our `average' cluster is 
around 100, and our fraction of detectable CVs for such a limiting magnitude is around
20 per cent, we should expect that they would have found around 20 CVs in their search,
which corresponds to an upper limit. 
This conclusion comes from the fact that we are considering 
an ideal model for the probability of detecting CVs through their variabilities,
 i.e. Fig. \ref{Fig06} shows the number/fraction  of predicted CVs
that should be detected in an ideal observation, which does not account for the 
realities of collecting observational data (e.g. distance, reddening, etc.).

In practice, some effects not considered here would lower
the fraction of CVs observed even further, such as

(i) GC are crowded fields;

(ii) large observational errors for faint stars;

(iii) WDs can have magnetic fields strong enough to avoid 
the instability in the disc (25 per cent, in the field);

(iv) the mass transfer rate computed here is a rough estimation. Thus,
some CVs could have smaller mass transfer rates and
be inactive instead of being DNe ones;

(v) the number of nights used to observe a specific cluster 
is less than the entire DN cycle;

(vi) heterogeneous observations unevenly distributed over time;

(vii) the estimate of luminosities is the highest possible 
in the DIM;

(viii) only normal outbursts are considered. Superoutbursts are present 
in short-period CVs as SU UMA and are more luminous, however the
recurrence time is much longer; this, in turn, would decrease the
probability.

Thus, after including all these complications, one should
have only a small fraction of the CVs in a GC detected
during the observations for a relatively low limiting magnitude such as
$\sim$ 26 mag. At least half of the predicted number would be expected.

\subsection{Future Observations}
\label{future observations}

From the last subsection, we realize that, in order to 
reach good detection rates through variability, one would have to observe the CVs during
 quiescence in very deep observations ($\gtrsim 27$ limiting magnitude) 
and in not too remote GCs ($\lesssim$ 5 kpc). For the typical limiting magnitudes
 achieved in previous surveys \citep[e.g.][]{Shara_2005,Pietrukowicz_2008} and
previous observations \citep[e.g.][]{Cohn_2010}, 
only a very small fraction of the CVs 
(considering that they are DNe) would be seen. 
So, it is not surprising that they detected only a small
number of CVs.

Fig. \ref{Fig06} is quite instructive and poses quite an interesting
result: {\it CVs in GCs are not, a priori, non-DNe}. In other words,
it is not easy to rule out the notion that most CVs in GCs are
DNe, especially by considering observational selection effects.
Even though the idea that CVs in GCs are preferentially magnetic has been
largely accepted and treated as coherent, from our results we cannot discard
the possibility that most of them are DNe, and due to observational selection
effects, we are not able to always detect them during
\citep[e.g.][]{Servillat_2011,Webb_2013}.

To sum up, any of the problems mentioned in Subsection \ref{probability} 
would lower the probability even further
and, in turn, not change the main conclusion here that
observational selection effects do change what one would expect to observe and
that the probability of detecting CVs in observations and campaigns is
extremely small.

Thus, what sort of observation should be performed in order
to observe the predicted CV populations in GCs?

\subsubsection{Variability during quiescence?}
\label{cv_observation_variability}

We could conclude in Subsection \ref{cv_dne} that, given the typical 
mass transfer rate of the predicted CVs, they should be DNe, unless
the WD magnetic field is strong enough to disrupt the inner part
of the disc. Additionally, we showed in Subsection \ref{probability} that
the chances of detecting a CV during its outburst are usually small, as
its intrinsic duty cycle is small.

One alternative that might help here would be the search for variabilities
during quiescence. With such an approach, whether a CV is in quiescence
or outburst can be overcome. However, reaching the large
magnitudes of faint CVs in GCs might be an issue. Nevertheless, 
the brightest CVs in a faint population should be detected
during deep observations. These correspond to the WZ Sge type of CVs.  

WZ Sge CVs have very low-mass donors
($\sim$ 0.08 ${\rm M_\odot}$) and their quiescent stages
last decades due to the extremely small mass transfer rate. Furthermore,
they are believed to be in the transition stage
between short-period and period bouncer CVs. Thus,
they are the prototype of the period bouncers of which 
they are more luminous.
Therefore, if there are, indeed, a lot of period bouncers in GCs,
the WZ Seg should reveal real trends associated with the CV
population in GCs.

This idea has already been proposed by \citet{Knigge_2012MMSAI},
and seems reasonable based on our results.

\subsubsection{X-rays}
\label{cv_observation_fuv}

Most of the predicted CVs in our models have X-ray luminosities between
$\sim 10^{29}$ and $\sim 10^{30}$ {\rm erg s$^{-1}$}. Although, a 9 per cent fraction
have $L_{\rm X} \lesssim 10^{29}$ {\rm erg s$^{-1}$}. As we have
already pointed out in Subsection \ref{cv_observation}, little effort
was used in order to recognize counterparts of X-ray sources
below $\sim 10^{32}$ erg s$^{-1}$. In this way, more effort should
be put forth in order to detect sources below $\sim 10^{30}$ erg s$^{-1}$
with secure optical counterparts. This might be difficult,
although doable.

Over and above, our results are consistent with what \citet{Byckling_2010} 
found for DNe within $\sim$ 200 pc in the solar neighbourhood. However, their 
sample is small and composed of bright X-ray DNe, which makes 
the comparison difficult. Nevertheless, if the CVs in GCs are predominantly
DNe, then we should be able to detect them in X-ray observations below $\sim 10^{30}$ 
erg s$^{-1}$.

We might say that looking deeper with {\it Chandra} may help to disentangle
the issue about the nature of CVs in GCs. On the one hand, the X-ray
counterparts are biased towards high X-ray luminosities which
implies an easier detection of magnetic CVs. Thus, deeper observations
would lead to the discovery of potential DNe. On the other hand,
the long recurrence time associated with CVs in GCs makes the
X-ray identification unlikely in the detection of such DNe,
since they live most of their lives in a quiescent stage.

\subsubsection{Other methods}
\label{cv_observation_other}

Once the faint X-ray sources are identified, the next step
should be to search for secure optical counterparts. In addition,
a combination of H$\alpha$ and FUV imaging with {\it HST} might
provide almost secure CVs, especially the WZ Sge population,
since they are brighter than typical period bouncers.

Alternatively, other approaches could help even more, like the
usage of quiescent late, and negative superhump, which are 
observed in some SU UMa stars. For instance,
\citet[][see their fig. 7]{Olech_2007} found high amplitude
($\sim$ 0.7 mag) negative superhumps in BF Ara.

In summary, a great deal of deep observations (combining different
approaches) might reveal the missing DNe population in GCs, 
and increase the knowledge about CVs in GCs as a whole. This set 
of deep observations yet to come might potentially solve the problem
concerning the observed lack of CVs in GCs relative to theoretical 
predictions.

\subsection{Spatial Distribution}
\label{spatial}

Another important property from an observational point of view
is the spatial distribution of the predicted CVs
and their properties with respect to their
positions inside the cluster.

Considering the six models, only $\sim$ 5 per cent of the 
predicted CVs are located in the core. Interestingly, most CVs
 are somewhere between the core and the half-mass radii.
This is easy to understand by considering either the relaxation 
process or the effect of dynamical interactions on CV progenitors.

Generally, there are three physical processes which shape the CV spatial distribution, 
namely, relaxation, dynamical interactions, and stellar/binary evolution. For simplicity,
let us start thinking of the CVs not affected by dynamics. For such cases, the initial mass segregation 
(connected with the relaxation process) for the most massive CV progenitors brings them close to the core.
After the CEP, the CVs might have masses greater or smaller than the average mass in the core.
If a CV has mass greater than the average mass in the core, then such a CV should migrate inwards,
into the core. On the other hand, if the CV has mass smaller than the average mass in the core, 
then the CV should be pushed outwards, away from the core. For less massive CV progenitors, the above-mentioned
process should be the same, although slower. 

CV progenitors are among the most massive stars, therefore the probability for dynamical interactions 
is effectively large, especially close to the core. For the CVs formed under the strong influence of dynamics, 
they are usually kicked out from the inner parts of the system due to interactions. Then, if the
CV progenitor is massive, it will sink to the core. Although it depends strongly
on the cluster evolution itself and if there is energy generation in the core by, for instance,
subsystems of black holes, intermediate-mass black holes, etc. In the case of large
energy generation in the core, the whole system expands (including the CVs), and so we should
not expect many CVs close to the core.

In general, the CV spatial distribution is influenced by relaxation, stellar
evolution and dynamical interactions, and drawing a general conclusion might be 
tricky since there are many issues to be considered in the analysis, like core
evolution, CV formation channel, CV formation time, CV initial position, among
others.

It is worth mentioning that the young (and bright) CVs 
in our models are more centrally distributed than the
older ones, although the spread is large (due to interactions
and different cluster properties). 

Our results are in good agreement with the spatial distribution
inferred by \citet{Cohn_2010}, particularly with respect to 
the magnitude of the CVs. \citet[][fig. 3]{Cohn_2010} show
a possible transitional path from the bright CVs to the faint
CVs. We also reproduce such a feature in our simulations, although
we have many more faint CVs in our data with respect to those
that they could detect. This is easily understood by considering
the observational limitations of their investigation ($\sim$ 26 mag) 
and the predominant magnitude of such faint CVs in our simulations
($\lesssim$ 27). 

\section{SUMMARY AND CONCLUSIONS}
\label{conclusion}

So far, we have exposed in this paper part of the first results obtained
from the analysis of six models simulated by MOCCA using the CATUABA code
as well as description of the code itself. 

The PDP of CVs in GC are in the last stage of their evolution! 
This, maybe, is the leading characteristic
of the CVs in GCs. In addition, if the WD magnetic field is irrelevant,
they all are DNe. 

In order to check if one could `easily' observe the DNe with typical
limiting magnitudes in previous surveys, a simple (and rather ideal) 
probability law was derived and computed, whose results are displayed in 
Fig. \ref{Fig06}. It shows clearly
that observational selection effects are hiding the real population of CVs in GCs.
This can also be thought of with respect to the CVs in the field 
\citep{Pretorius_2007}.

In order to observe the population of CVs in GCs, one should
be able to perform deeper observations in such a way that the search
would be during quiescence. Something that was already
noticed by \citet{Knigge_2012MMSAI} and, to some extent, done 
by \citet{Cohn_2010}.

Interestingly, some alternatives exist in order to look for CVs during
quiescence. For instance, \citet{Cohn_2010} used H${\rm \alpha}$ imaging
(which allows a search for short-period variability) in combination with
optical observations of {\it Chandra} X-ray sources. 
Other ways that might help are FUV with {\it HST} \citep{Knigge_2003}
and quiescent negative superhump \citep{Olech_2007}.

In any event, we defend here that the CVs in GCs are not necessarily magnetic. 
The arguments for magnetic CVs seem to be biased in addition to
neglecting some other issues, like observational selection effects and 
incompleteness in surveys.

As has already pointed out by \citet{Knigge_2012MMSAI},
the natural solution to this problem is a survey for DNe in GCs 
that guarantees the detection of at least a few WZ Sge systems.

More about the CV populations like the progenitor population, 
the formation-age population, and also the evolutionary
properties will be discussed, as was said before, in a separate
work. Additionally, the apparent problem entailed with the Kroupa
IBP is also left to another paper since more models will be needed
and an attempt to overcome such an inconsistency will be discussed.

Thus, for this paper, the results displayed in Section \ref{results}
 can be summarized as follows.

(i) GCs with Kroupa IBP have no CVs formed purely via binary stellar evolution, 
i.e. dynamics have to act in order to produce CVs. The reason can be either
the efficiency of the feeding algorithm or the high CEP efficiency
adopted in BSE.

(ii) The main mechanism in CV formation for the Kroupa models is 
exchange ($\sim \; 2/3$), which is not true for other models (Standard models) 
in which the main mechanism is the binary stellar evolution.

(iii) Almost all CVs in the present-day clusters are formed below the gap. Additionally, most of the CVs at 12 Gyr 
($\sim$ 87 per cent) are period bouncers with BD as donors, having periods between 1.5 and 3h.

(iv) We obtained, as in previous simulations, a population of CVs with WDs with small
 masses ($\sim$ 0.3 ${\rm M_\odot}$). Moreover, in our case, the massive WDs ($\sim$ 0.7 ${\rm M_\odot}$) 
 correspond mainly to CVs formed because of
strong dynamical interactions (mainly exchanges), which is in good agreement 
with previous works \citep[e.g.][]{Ivanova_2006}.

(v) All CVs are DNe if there is no strong WD magnetic field such that it can disrupt
the inner part of the disc.

(vi) The duration of the outburst varies from 1-10 days and the recurrence time 
(duration of the quiescent phase) varies from 10-10000 days. This implies that the 
probability of detecting most of the CVs during outburst is extremely low, even considering an 
ideal situation in which all nights -- during the DN cycle -- are observed.

(vii) During quiescence, the eclipsing CVs would be detected in a very deep observation
(apparent visual magnitude $\gtrsim 27$ ) in a very close cluster ($\lesssim 5$ kpc).

(viii) GCs are old objects which implies that the population of CV tends to be older than 
the observed population in the field. Thus, at the present-day (12 Gyr), the population
 of CVs is dominated by low-mass donors. In other words, the properties of the population 
of CV changes with time through the cluster evolution and this has to be taken into account
for meaningful comparisons between observations and predictions.

(ix) in order to solve the problem regarding the nature of CVs in GCs, 
additional effort should be put into the optical identification of faint {\it Chandra} X-ray
sources ($\lesssim 10^{30}$ erg s$^{-1}$) and the utilization of other techniques (e.g. H$\alpha$ 
and FUV imaging with {\it HST}, quiescent negative superhumps) combined together
with the aim of detecting almost guaranteed DNe, specially because any conclusions drawn from a 
comparison between the results of our simulations and observations of CVs with small X-ray 
luminosities should be taken with a grain of salt, since the observational sample can be regarded 
as something of an upper limit, due to an increased probability of contamination from active 
binaries, chromospherically active stars, accreting NSs and BHs, etc.

(x) the best region inside GCs to search for CVs should be between the core and
half-mass radii.

Finally, some comments are needed concerning the Monte Carlo approach
and the population synthesis code utilized in this
work.

In Section \ref{nbody_codes} we briefly described some comparisons
made between MOCCA and NBODY6 that have led to good agreement 
between the two. Therefore, we think the approach chosen in this
work is reasonable and efficient since the MOCCA code is significantly
faster than the most advanced version of NBODY6++GPU.

Even then, the main limitation turns out to be related to
the binary stellar evolution code. Some limitations
with regard to the CV evolution were commented on in Section 
\ref{bse_code}, like the absence of expansion of the donor star for
long-period CVs and the old scaling for the angular momentum
losses above and below the gap. Nevertheless, from a statistical
point of view, such improvements should not change the main
conclusions drawn in this work.

\section{PERSPECTIVES}
\label{perspectives}

The first results of this initial investigation on CVs in GCs are very 
interesting, especially those associated with observational selection 
effects and the absence of CVs when using Kroupa IBP. CVs are a potentially 
useful tracer population for constraining primordial binary distributions
as we know the physical mechanisms that drive CV formation and evolution. 
The IBP can be modified and constrained using CVs by reproducing their 
observed properties in simulation models.

Nevertheless, this work is just the beginning.
Since MOCCA is extremely fast in comparison with N-body codes, several hundreds
of models can be simulated (and are being simulated) in order to develop
a statistical basis the for a constraint of the
overall population of CVs in GCs, the correlation between their properties, and the
clusters properties, etc.

Naturally, investigations not only focused on CVs can and should be done.
It is also worthwhile to study CV siblings like AM CVn and symbiotic stars.

Future works will concentrate on the extension of the CATUABA code with
the aim of also analysing AM CVn and symbiotic stars. After extending this
code, the same six models used here will be used in order to check consistencies
and coherence in a similar way done in this initial investigation on CVs.

Once the CATUABA code is fully automated and capable of studying CVs, AM CVns
and symbiotic stars in GCs simulated by MOCCA, we intend to apply it to the examination
of the MOCCA-SURVEY database that is being created in the PSK cluster at CAMK. We expect
that such an approach will represent a powerful tool in the analysis of particular
objects in GCs since the cluster parameter space covered by this survey is huge.

To sum up, as was said, this is the first paper of a series that has the aim
of contributing to the most important open questions regarding such
fascinating objects and globular clusters.

\section*{Acknowledgements}

We would like to kindly thank J{\'o}zef Smak, Arkadiusz Olech and Wojtek Pych
for useful discussions and suggestions, which
made the quality of the paper increase substantially. We would also like
to thank the anonymous referee for the numerous comments that improved this 
work significantly, especially its presentation.
DB is in debt to J{\'o}zef Smak and Arkadiusz Olech for pleasantly explaining 
many issues associated with the theory concerning cataclysmic variables.
DB was supported by the CAPES foundation, Brazilian Ministry of Education
through the grant BEX 13514/13-0. MG, AH and AA were supported by
Polish Ministry of Science and Higher Education, and by the National
Science Centre through the grants DEC-2012/07/B/ST9/04412 and
DEC-2011/01/N/ST9/06000. AA would also like to acknowledge support
from Polish Ministry of Sciences and Higher Education through the
grant DEC-2015/17/N/ST9/02573 and partial support from Nicolaus
Copernicus Astronomical Centre's grant for young researchers. 

\bibliographystyle{mnras}
\bibliography{references}

\appendix

\section[]{Abbreviations}
\label{ap1}

In this appendix, we define all the abbreviation used
in this work in order to provide the reader with an easy
way to recognize them instead of going back through the
text.

\begin{description}
\NumTabs{7}
\item[BD] \tab{brown dwarf}
\item[BSE] \tab{binary stellar evolution}
\item[CEP] \tab{common envelope phase}
\item[CV] \tab{cataclysmic variable}
\item[DIM] \tab{disc instability model}
\item[DNe] \tab{dwarf novae}
\item[GC] \tab{globular cluster}
\item[IBP] \tab{initial binary population}
\item[IMF] \tab{initial mass function}
\item[MS] \tab{main sequence}
\item[PDP] \tab{present-day population}
\item[SDI] \tab{strong dynamical interaction}
\item[WD] \tab{white dwarf}
\item[WDI] \tab{weak dynamical interaction}
\end{description}

\bsp

\label{lastpage}

\end{document}